\documentclass[twocolumn,english,aps,prd,showpacs]{revtex4}
\usepackage[T1]{fontenc}
\usepackage[latin9]{inputenc}
\usepackage{color}
\usepackage{amsmath}
\usepackage{graphicx}
\usepackage{amssymb}
\usepackage{esint}
\usepackage{babel}

\makeatletter

\@ifundefined{textcolor}{}
{%
 \definecolor{BLACK}{gray}{0}
 \definecolor{WHITE}{gray}{1}
 \definecolor{RED}{rgb}{1,0,0}
 \definecolor{GREEN}{rgb}{0,1,0}
 \definecolor{BLUE}{rgb}{0,0,1}
 \definecolor{CYAN}{cmyk}{1,0,0,0}
 \definecolor{MAGENTA}{cmyk}{0,1,0,0}
 \definecolor{YELLOW}{cmyk}{0,0,1,0}
 }

\makeatletter

\newcommand{\CA}{{\cal A}}

\newcommand{\CD}{{\cal D}}
\newcommand{\CE}{{\cal E}}
\newcommand{\CF}{{\cal F}}

\newcommand{\CQ}{{\cal Q}}
\newcommand{\CR}{{\cal R}}

\newcommand{\CM}{{\cal M}}
\newcommand{\average}[1]{\left\langle #1 \right\rangle_\CD}

\newcommand{\averageM}[1]{\left\langle #1 \right\rangle _{{\cal M}}}
\newcommand{\averageE}[1]{\left\langle #1 \right\rangle _{{\cal E}}}

\newcommand{\averageF}[1]{\left\langle #1\right\rangle _{\CF}}

\newcommand{\averageFi}[1]{\left\langle #1\right\rangle _{\CF_{\ell}}}

\makeatother

\begin{document}

\title{Multiscale cosmology and structure--emerging Dark Energy:\\
A plausibility analysis}

\author{Alexander Wiegand$^{1,2,3}$ and Thomas Buchert$^{3}$}

\affiliation{$^{1}$Fakult\"at f\"ur Physik, Universit\"at Bielefeld, Universit\"atsstra\ss e~25, D--33615 Bielefeld, Germany}

\affiliation{$^{2}$Institut f\"ur Theoretische Physik, KIT, Campus S\"ud, Wolfgang--Gaede--Str.~1, D--76131 Karlsruhe, Germany}

\affiliation{$^{3}$Universit\'e Lyon~1, Centre de Recherche Astrophysique de Lyon,
CNRS UMR 5574, 9 avenue Charles Andr\'e, F--69230 Saint--Genis--Laval,
France\\ \\
 Emails: wiegand@physik.uni--bielefeld.de and buchert@obs.univ--lyon.fr}



\begin{abstract}
Cosmological backreaction suggests a link between structure formation and the expansion history of the Universe. In order to quantitatively examine this connection, we dynamically investigate a volume partition of the Universe into over-- and underdense regions. This allows us to trace structure formation using the volume fraction of the overdense regions $\lambda_{\CM}$ as its characterizing parameter. Employing results from cosmological perturbation theory and extrapolating the leading mode into the nonlinear regime, we construct a three--parameter model for the effective cosmic expansion history, involving $\lambda_{\CM_{0}}$, the matter density $\Omega_{m}^{\CD_{0}}$, and the Hubble rate $H_{\CD_{0}}$ of today's Universe. Taking standard values for $\Omega_{m}^{\CD_{0}}$ and $H_{\CD_{0}}$ as well as a reasonable value for $\lambda_{\CM_{0}}$, that we derive from $N$--body simulations, we determine the corresponding amounts of backreaction and spatial curvature.
We find that the obtained values that are sufficient to generate today's structure also lead to a $\Lambda$CDM--like behavior of the scale factor, parametrized by the same parameters $\Omega_{m}^{\CD_{0}}$ and $H_{\CD_{0}}$, but without a cosmological constant. However, the temporal behavior of $\lambda_{\CM}$ does not faithfully reproduce the structure formation history.
Surprisingly, however, the model matches with structure formation with the assumption of a low matter content, $\Omega_{m}^{\CD_{0}}\approx3\%$, a result that hints to a different interpretation of part of the backreaction effect as kinematical Dark Matter.

A complementary investigation assumes the $\Lambda$CDM fit--model for the evolution of the global scale factor by imposing a global replacement of the cosmological constant through backreaction, and also supposes that a Newtonian simulation of structure formation provides the correct volume partition into over-- and underdense regions. From these assumptions we derive the corresponding evolution laws for backreaction and spatial curvature on the partitioned domains. We find the correct scaling limit predicted by perturbation theory, which allows us to rederive higher--order results from perturbation theory on the evolution laws for backreaction and curvature analytically. This strong backreaction scenario can explain structure formation and Dark Energy simultaneously.
 
We conclude that these results represent a conceptually appealing approach towards a solution of the Dark Energy and coincidence problems. Open problems are the still too large amplitude of initial perturbations that are required for the scenarios proposed, and the role of Dark Matter that may be partially taken by backreaction effects. Both drawbacks point to the need of a reinterpretation of observational data in the new framework.
\end{abstract}
\maketitle

\section{Introduction}

One decade has passed since the first long distance SN measurements (see \cite{SNIa:Union,SNIa:Constitution,SNIa:Essence} for the latest data)
led to the belief in an accelerating expansion of our Universe and
the postulation of Dark Energy (even
if the effect is model dependent \cite{bolejkoandersson,huntsarkar} and there
are some possibilities that do not need acceleration \cite{LTB:review,celerier}). 
There is still no convincing theoretical understanding
of the experimental result, and the creativity of theoretical physicists
has led to a plethora of proposals on the nature of this new energy component, see e.g. \cite{DE:pilar}.

In the same decade, however, initially independent of the development
in the Dark Energy sector, our knowledge about another, long-standing
problem in relativistic cosmology has grown to an extent that allowed
to connect it to the Dark Energy problem: the question of how effectively
inhomogeneities in matter and geometry would influence the global behavior of the Universe \cite{ellis:average}.
As general relativity is far too complex
to simply solve the Einstein equations for an inhomogeneous matter
distribution and calculate the effect through tensorial averaging techniques, there have been several 
approaches to get a handle on this problem \cite{ellisbuchert}. The one which lies at the basis of
this paper was introduced in \cite{buchert:dust} and
uses a truncated hierarchy of the averaged scalar parts of Einstein's equations. 
As will be recalled in Section~\ref{sub:The-average-Einstein}, this strategy reduces the problem
to a set of three coupled equations for four independent variables characterizing
the state of some spatial domain in the Universe, its volume, mean
density, mean scalar curvature and backreaction, linked to
the degree of inhomogeneity of the domain's deformation.
For details see Sec. \ref{sub:The-average-Einstein} and \cite{buchert:dust,buchert:fluid,buchert:jgrg,buchert:review}.

Based on this framework the connection of the two above-mentioned problems was made in \cite{rasanen:de}
and \cite{kolb:backreaction}, where the authors
proposed that the backreaction of the inhomogeneities developing in
the era of structure formation could lead to the effect interpreted
as accelerated expansion. This would also solve the coincidence problem,
i.e. answer the question why the acceleration takes place at a moment
of the evolution when the Universe leaves its homogeneous initial
state. For this and other more theoretically founded reasons, there
have already been some applications of the formalism of the averaged
equations to structure formation. One has been carried out by R\"as\"anen
\cite{rasanen:peakmodel} (generalizing his earlier work on a two--scale model \cite{rasanen:acceleration}). 
He used a model of structure formation
in which the extreme points of the initial overdensity field are evolved
separately using a spherical LTB model. His treatment did not give
evidence for an accelerated expansion, but he could identify a signal
of onset of an eventually strong backreaction regime at the right time. Another study that also follows
a multiscale approach as in the present paper has been performed
by Wiltshire \cite{wiltshire:clocks,wiltshire:avsolution}. He separated
the universe model into {}``voids'' and {}``walls'' and
studied the backreaction effects resulting from fluctuations due to this partitioning.
To solve the equations he assumed the backreaction term to vanish
on each of the subregions and set the curvature on the overdense regions
to zero (as in the two--scale model of R\"as\"anen \cite{rasanen:acceleration}). For the underdense regions he considered a Friedmann--like
constant curvature term. Under these assumptions he also did not find
volume acceleration, but described an effect that should result from
the different lapse of time in underdense void regions and overdense
matter dominated regions. This allowed him to fit his model to a fair number of
data \cite{wiltshire:clocks,wiltshire:obs,wiltshire:timescape}, despite his assumption that the backreaction effect on the individual
domains was absent. We shall come back to his approach later. Other interesting considerations about a multiscale approach
to self--gravitating systems may be found in \cite{multi}.

We wish to take one step back and return to the more general problem
(i.e. without using R\"as\"anen's or Wiltshire's assumptions), by presenting the general
framework for separations of the averaged equations into subdomains
of the Universe. This has been begun in \cite{buchertcarfora} by 
investigating a general volume partition of cosmological hypersurfaces.
We will show that this separation is consistently possible also for the
evolution equations. The advantage of not neglecting any effect is
impaired by the caveat not to be able to close the system of averaged equations
without further assumptions on evolution laws for backreaction and curvature. 
We are going to investigate a volume partition 
of the Universe into initially overdense $\CM$--regions
and underdense $\CE$--regions in analogy with Wiltshire's approach. However,
instead of setting the backreaction on $\CM$ and $\CE$ to zero,
we will first use perturbative results on the early evolution of backreaction \cite{bks}, \cite{gaugeinv,li:scale,li:thesis}:
in perturbation theory one finds that the backreaction
term on a subdomain of a matter dominated universe model decays only as
$a^{-1}$ with the scale factor (a result found in various papers in relativistic cosmology \cite{rasanen:de,kolb:backreaction,gaugeinv,li:scale,li:thesis} that confirms the formally analogous situation for the backreaction term in Newtonian cosmology \cite{bks}). The latter work also shows that it is the leading term in a nonperturbative approximation based on the exact averaged equations and the Zel'dovich approximation \cite{zeldovich,buchert89,buchert92,buchert93} for the fluctuations. Furthermore, this work selects the leading term as the dominant contribution on large, ``quasilinear" scales. 
Note here that in some work this leading mode is dismissed due to the property that its coefficient function is a full divergence and must hence vanish by imposing periodic boundary conditions. While this latter remark is true \cite{buchertehlers}, the mode is nevertheless there in the interior of a periodic Newtonian or quasi--Newtonian simulation box. Moreover, in a proper relativistic theory these terms are not full divergences and there are no boundary conditions to be satisfied apart from the constraints on the initial hypersurface.  

As a result of this behavior of the leading mode the importance of the backreaction term, although being a second--order quantity, quickly rises with
respect to the matter density which goes as $a^{-3}$. This perturbative mode falls on an exact scaling solution \cite{morphon} of the general problem and we, hence, consider 
evolution laws for backreaction and curvature by extrapolation of the 
perturbative mode along this particular scaling solution. We emphasize that this extrapolation of the leading perturbative mode already furnishes a nonperturbative model due to the fact that the mode is referred to an evolving background that includes backreaction \cite{buchert:darkenergy,buchert:static}, \cite{kolb:backgrounds}. As an aside we here note that a perturbation theory on the evolving background given by the averaged equations has not yet been concisely investigated; forthcoming work will especially focus on the formulation of such a theory, which will be key to an adequate generalization of perturbation theories in the averaging framework.

Our second approach uses the general formalism and is specialized to a concrete
model by making use of $N$--body simulation data and the assumption that
backreaction globally acts like a cosmological constant. This latter approach illustrates
a strong backreaction scenario that simultaneously describes
accelerated expansion and the structure formation history. $N$--body simulation data have recently also been used to estimate the backreaction in a quasi--Newtonian setting through an estimation of the post--Newtonian potential \cite{zhao}. We here allow for a more general reinterpretation of Newtonian simulations that -- combined with the relativistic equations -- takes scalar curvature into account.

This paper is organized as follows: Section~\ref{sec:Volume-partitions}
shortly reviews the averaged equations and the occurrence of the backreaction
terms. Then the relevant separation formulae are presented and the
consistency of the resulting evolution equations is checked. The question of
how accelerated expansion may be understood in the context of these equations is addressed
in Subsection~\ref{sub:A-first-example}. Section~\ref{sec:A-simple-scaling}
uses the assumption of the $a^{-1}$--scaling of the backreaction terms
to clarify the connection between structures today, the initial density
fluctuations and the accelerated expansion. The comparison of \ref{sub:Comparison}
shows that the specific model is not sufficient to explain structure
formation and points out why this could also not have been
expected. The discussion will concentrate on how one has to devise
the evolution of inhomogeneities in the context of the averaged equations
in order to fully explain the phenomenon of Dark Energy.

Section~\ref{sec:Modelling-structure-formation} investigates how
the perturbative $a^{-1}$--behavior has to be extended to combine
structure formation and accelerated expansion. 
In \ref{sec:Power-series}A we motivate the choice of the $a^{-1}$--scaling
for the backreaction term on $\CM$ and $\CE$ and rederive the results
of \cite{bks}, \cite{li:scale,li:thesis} in our context. We also perform
a quantitative estimation of the amount of initial backreaction necessary
for structure formation. Finally, we give quantitative conclusions in  \ref{sec:Power-series}B,
and discuss in Sec. \ref{sec:Discussion}
what extensions could be made to the models considered and comment
on the need for a substantial reinterpretation of observational data.

\section{Volume partitions of the average universe\label{sec:Volume-partitions}}

\subsection{The averaged Einstein equations for compact spatial domains\label{sub:The-average-Einstein}}

The basis for our investigation of the link of structure formation
to the expansion history of the Universe is to consider integral properties
of scalar variables in a pressureless -- i.e. dust -- universe model, given in \cite{buchert:dust}, 
later refined to be applicable to perfect fluid matter models in \cite{buchert:fluid}, where
also compact formulations of the averaged equations of  \cite{buchert:dust} can be found. 
The space time is foliated into flow--orthogonal hypersurfaces featuring the line--element
\begin{equation}
ds^{2}=-dt^{2}+g_{ij}dX^{i}dX^{j}\;,
\end{equation}
where the proper time $t$ labels the hypersurfaces and $X^i$ are Gaussian normal coordinates
(locating free--falling fluid elements or generalized fundamental observers)
in the hypersurfaces. $g_{ij}$ is the
full inhomogeneous three--metric of the hypersurfaces of constant proper
time. Such a split corresponds in the ADM formalism
to the choices of a constant lapse and a vanishing shift. Extensions of this formalism to nonvanishing shift and
tilted foliations have been given in \cite{larena} and \cite{brown1}, in \cite{brown2} with applications to perturbation theory.
For considerations of covariance and gauge issues see \cite{veneziano2}.

On these hypersurfaces we want to study the evolution of compact
spatial domains $\CD$, comoving with the fluid. This latter property ensures that the domain is frozen into
the general three--metric, i.e. its shape encodes the geometrical structure of the local inhomogeneities.
Note that in Newtonian spacetimes it can be shown that the morphology of the boundary of a comoving domain already contains
comprehensive information of the matter distribution, including all higher--order correlations. The morphology--characterizing set of
integral geometrical measures is known as the Minkowski functionals, from which the
backreaction term can be built (see \cite{buchert:review} for details).
One fundamental quantity characterizing such a domain is its volume, which is the only such measure used here, since we
wish to address questions related to the size of the domains and its time--derivatives only,
\begin{equation}
\left|\CD\right|_{g}:=\int_{\CD}\, d\mu_{g} \;,
\end{equation}
where $d\mu_{g}:=\sqrt{\,^{\left(3\right)}g\left(t,X^1, X^2, X^3\right)}dX^{1}dX^{2}dX^{3}$.
From the domain's volume one may define a scale factor
\begin{equation}
a_{\CD}\left(t\right):=\left(\frac{\left|\CD\right|_{g}}{\left|\CD_{i}\right|_{g}}\right)^{\frac{1}{3}}\;,
\label{eq:scale-factor}
\end{equation}
encoding the average stretch of all directions of the domain. For
wild changes of the shape of the initial domain $\CD_{i}$ one might
want to know more about the evolution of other morphological characteristics 
to deduce directional expansion information, and would therefore have to extend the analysis.

Concentrating on the volume and the effective scale factor alone, one can derive, from the Einstein equations
with a pressureless fluid source,  the
following set of equations governing its evolution:
\begin{eqnarray}
3\frac{\ddot{a}_{\CD}}{a_{\CD}} & = & -4\pi G\average{\varrho}+\CQ_{\CD}+\Lambda
\label{eq:Raychaudhuri-Mittel}\\
3H_{\CD}^{2} & = & 8\pi G\average{\varrho}-\frac{1}{2}\average{\CR}-\frac{1}{2}\CQ_{\CD}+\Lambda
\label{eq:Hamilton-Mittel}\\
0 & = & \partial_{t}\average{\varrho}+3H_{\CD}\average{\varrho}\;,
\label{eq:Konservation-Mittel}
\end{eqnarray}
where the average over scalar quantities is defined as
\begin{equation}
\left\langle f\right\rangle _{\CD}\left(t\right):=\frac{\int_{\CD}\, f\left(t,X^1,X^2,X^3\right)d\mu_{g}}{\int_{\CD}\, d\mu_{g}}\;,
\label{eq:Def-Mittel}
\end{equation}
and where $\varrho$, $\CR$ and $H_{\CD}$ denote the local matter
density, the Ricci scalar of the three--metric
$g_{ij}$, and the domain dependent Hubble rate $H_{\CD}:=\dot{a}_{\CD} / a_{\CD}$,
respectively. The kinematical backreaction $\CQ_{\CD}$ is defined as
\begin{equation}
\CQ_{\CD}:=\frac{2}{3}\left(\average{\theta^{2}}-\average{\theta}^{2}\right)-2\average{\sigma^{2}}\;,
\label{eq:Def-Q}
\end{equation}
where $\theta$ is the local expansion rate and $\sigma^{2}:=1/2 \sigma_{ij}\sigma^{ij}$
is the squared rate of shear. Note that $H_{\CD}$ is defined as $H_{\CD}=1/3\average{\theta}$.
$\CQ_{\CD}$ is composed of the
variance of the local expansion rates, $\average{\theta^{2}}-\average{\theta}^{2}$,
and the averaged shear scalar $\average{\sigma^{2}}$ on the domain
under consideration. For a homogeneous domain it is zero.
It therefore encodes the departure from homogeneity and is supposed
to be particularly important in the late, inhomogeneous phase of the
Universe at the epoch of structure formation.

The integrability condition connecting Eqs.~(\ref{eq:Raychaudhuri-Mittel})
and (\ref{eq:Hamilton-Mittel}) reads
\begin{equation}
a_{\CD}^{-2}\partial_{t}\left(a_{\CD}^{2}\langle\CR\rangle_{{\cal \CD}}\right)=-a_{\CD}^{-6}\partial_{t}\left(a_{\CD}^{6}\CQ_{\CD}\right)\;,
\label{eq:Integrability-curv}
\end{equation}
which already shows an important feature of the averaged equations
as they in general couple the evolution of the backreaction term, and hence extrinsic
curvature inhomogeneities (or in this picture matter inhomogeneities), to the average intrinsic curvature. Unlike in the case
of a standard Friedmannian evolution the curvature term here is not
restricted to an $a_{\CD}^{-2}$ scaling behavior but is dynamical in
the sense that it may be an arbitrary function of $a_{\CD}$. We emphasize that 
the essential effect of backreaction models is not a large magnitude of $\CQ_{\CD}$, but a 
dynamical coupling of a nonvanishing $\CQ_{\CD}$ to the averaged scalar curvature, changing the temporal
behavior of this latter.

One may also define cosmological parameters on the domain investigated.
In analogy to the Friedmannian case they are derived from the Hamilton
constraint (\ref{eq:Hamilton-Mittel}) by a division by $3H_{\CD}^{2}$.
For different spatial domains, which
will be denoted by $\CD$, $\CM$ and $\CE$, we formulate the cosmological
parameters in a generic way by taking $\CF$ to denote one of the domains
$\CD$, $\CM$, $\CE$ (this will hold for the whole text, but depending
on the context $\CF$ may also only denote $\CM$ and $\CE$). The
definitions are 
\begin{eqnarray}
\Omega_{m}^{\CF} & := & \frac{8\pi G}{3H_{\CD}^{2}}\langle\varrho\rangle_{\CF}\;\;;\;\;\Omega_{\Lambda}^{\CF}:=\frac{\Lambda}{3H_{\CD}^{2}}
\label{eq:Def-Omega-Parameter}\\
\Omega_{\CR}^{\CF} & := & -\frac{\langle\CR\rangle_{\CF}}{6H_{\CD}^{2}}\;\;;\;\;\Omega_{\CQ}^{\CF}:=-\frac{\CQ_{\CF}}{6H_{\CD}^{2}}\;,\nonumber 
\end{eqnarray}
where the decision to divide by $H_{\CD}^{2}$ for every $\CF$ will
become clear from the definition of $\CM$, $\CE$ and $\CD$ in the
next section. Using those definitions, the Hamilton constraint (\ref{eq:Hamilton-Mittel})
for a domain $\CF$ reads 
\begin{equation}
\Omega_{m}^{\CF}+\Omega_{\Lambda}^{\CF}+\Omega_{\CR}^{\CF}+\Omega_{\CQ}^{\CF}=\frac{H_{\CF}^{2}}{H_{\CD}^{2}}\;,
\label{eq:HamiltonDparameter}
\end{equation}
which means that the dimensionless parameters $\Omega$ only add up
to $1$ for the domain $\CD$. On other domains they may add up to
a more or less arbitrary positive value, depending on whether the
corresponding region $\CF$ expands faster or slower than the $\CD$
region.

To point out the analogy with the $k=0$ Friedmann equations one may recast
(\ref{eq:Raychaudhuri-Mittel})-(\ref{eq:Konservation-Mittel}) combining
the backreaction and the curvature terms to an effective density and
pressure: 
\begin{eqnarray}
3\frac{{\ddot{a}}_{\CD}}{a_{\CD}} & = & \Lambda-4\pi G(\varrho_{\rm eff}^{\CD}+3{p}_{\rm eff}^{\CD})\;\;;\;\;3H_{\CD}^{2}=\Lambda+8\pi G\varrho_{\rm eff}^{\CD}\nonumber \\
0 & = & {\dot{\varrho}}_{\rm eff}^{\CD}+3H_{\CD}\left(\varrho_{\rm eff}^{\CD}+{p}_{\rm eff}^{\CD}\right)\;,
\label{eq:effectivefriedmann}
\end{eqnarray}
where the effective densities are defined as 
\begin{eqnarray}
\varrho_{{\rm eff}}^{{\CD}} & := & \average{\varrho}-\frac{1}{16\pi G}{\CQ}_{{\CD}}-\frac{1}{16\pi G}\average{{\cal R}}
\label{eq:equationofstate}\\
{p}_{{\rm eff}}^{{\CD}} & := & -\frac{1}{16\pi G}{\CQ}_{{\CD}}+\frac{1}{48\pi G}\average{{\cal R}}\;.\nonumber 
\end{eqnarray}
In this sense, $\CQ_{\CD}$ and $\average{\CR}$ may be combined to
some kind of dark fluid component that is commonly referred to as $X$--matter.
One quantity characterizing this $X$--matter is its
equation of state
\begin{equation}
w_{\Lambda,\rm{eff}}^{\CD}:=\frac{{p}_{{\rm eff}}^{{\CD}}}{\varrho_{{\rm eff}}^{{\CD}}-\average{\varrho}}=\frac{\Omega_{\CQ}^{\CD}-\frac{1}{3}\Omega_{\CR}^{\CD}}{\Omega_{\CQ}^{\CD}+\Omega_{\CR}^{\CD}}\;,
\label{eq:eq-of-state}
\end{equation}
which is an effective one due to the fact that backreaction and
curvature give rise to an \emph{effective} energy density and pressure.

\subsection{Separation formulas for arbitrary partitions\label{sub:Separation-formulae}}

After the short review of the averaging formalism we present in the following how the resulting
equations can be separated, if one wants to consider the behavior of
subdomains of some {}``global'' region $\CD$, which we may assume to be associated with a
(postulated) scale of homogeneity. We consider,
a priori, arbitrary partitions of the spatial hypersurfaces into subregions
$\CF_{\ell}$, which themselves consist of elementary space entities
$\CF_{\ell}^{\left(\alpha\right)}$ that may be associated with some
averaging length scale. The idea will be to choose entities
$\CF_{\ell}^{\left(\alpha\right)}$ in a way that they share some common properties. In our case this
property will be the value of the overdensity field, but the separation
is general and also valid for other choices. To be more precise we
want to have $\CD=\cup_{\ell}\CF_{\ell}$ where $\CF_{\ell}:=\cup_{\alpha}\CF_{\ell}^{\left(\alpha\right)}$
and $\CF_{\ell}^{\left(\alpha\right)}\cap\CF_{m}^{\left(\beta\right)}=\emptyset$
for all $\alpha\ne\beta$ and $\ell\ne m$.
In the sequel we follow closely the investigation in \cite{buchertcarfora} of a general volume partitioning of cosmological hypersurfaces.

The average of the scalar valued function $f$ on the domain $\CD$,
\begin{equation}
\langle f\rangle_{\CD}:=|{\CD}|_{g}^{-1}\int_{{\CD}}f\, d\mu_{g}\;,
\end{equation}
defined in (\ref{eq:Def-Mittel}), where $|{\CD}|_{g}$ denotes
its volume $|{\CD}|_{g}:=\int_{{\CD}}d\mu_{g}$,
may then be split into the averages of $f$ on the subregions $\CF_{\ell}$,
in the form
\begin{eqnarray}
\average f & = & \sum_{\ell}|{\CD}|_{g}^{-1}\sum_{\alpha}\int_{\CF_{\ell}^{(\alpha)}}f\, d\mu_{g}
\label{eq:split}\\
 & = & \sum_{\ell}\frac{\left|\CF_{\ell}\right|_{g}}{\left|\CD\right|_{g}}\averageFi f=\sum_{\ell}\lambda_{\ell}\averageFi f \;,\nonumber 
\end{eqnarray}
where we have introduced 
\begin{equation}
\lambda_{\ell}:=\frac{\left|\CF_{\ell}\right|_{g}}{\left|\CD\right|_{g}}\;,
\end{equation}
i.e. the volume fraction $\lambda_{\ell}$ of the subregion $\CF_{\ell}$.
Regarding the dynamical equations we wish to separate, i.e. (\ref{eq:Raychaudhuri-Mittel}),
(\ref{eq:Hamilton-Mittel}) and (\ref{eq:Integrability-curv}), Equation
(\ref{eq:split}) directly provides the expressions for the scalar
functions $\varrho$, $\CR$ and $H_{\CD}:=1 / 3\average{\theta}$.
Only $\CQ_{\CD}$ as defined in (\ref{eq:Def-Q}) does not split in this simple way
due to the $\average{\theta}^{2}$--term. A short calculation rather
provides
\begin{equation}
\CQ_{\CD}=\sum_{\ell}\lambda_{\ell}\CQ_{\ell}+3\sum_{\ell\neq m}\lambda_{\ell}\lambda_{m}\left(H_{\ell}-H_{m}\right)^{2}
\label{eq:Q-split}
\end{equation}
as the correct formula. $\CQ_{\ell}$ is defined as in (\ref{eq:Def-Q})
with $\CF_{\ell}$ instead of $\CD$. The shear part $\averageFi{\sigma^{2}}$
is completely absorbed in $\CQ_{\ell}$, whereas the variance of the
local expansion rates, $\average{\theta^{2}}-\average{\theta}^{2}$,
is partly contained in $\CQ_{\ell}$ but also generates the extra term
$3\sum_{\ell\neq m}\lambda_{\ell}\lambda_{m}\left(H_{\ell}-H_{m}\right)^{2}$.
This is because the part of the variance that is present in $\CQ_{\ell}$,
namely $\averageFi{\theta^{2}}-\averageFi{\theta}^{2}$, only takes
into account points inside $\CF_{\ell}$. To restore the variance
that comes from combining points of $\CF_{\ell}$ with others in $\CF_{m}$,
the extra term containing the averaged Hubble rates emerges.

One may also define a scale factor $a_{\ell}$ analogous to (\ref{eq:scale-factor})
for each of the subregions $\CF_{\ell}$. As their definition implies
that they are disjoint, it follows that $\left|\CD\right|_{g}=\sum_{\ell}\left|\CF_{\ell}\right|_{g}$
and we may therefore define $a_{\CD}^{3}=\sum_{\ell}a_{\ell}^{3}$.
An important detail when using the equation in this form is that,
at the initial time when $a_{\CD}$ is equal to one, the scale factors $a_{\ell}$
of the subregions will not be equal to one as well. This
different normalization has to be taken into account when scaling
the local $\CM$ and $\CE$ parameters.

Differentiating $a_{\CD}^{3}=\sum_{\ell}a_{\ell}^{3}$ twice with
respect to the foliation time and using the result for $\dot{a}_{\ell}$,
we finally provide the relation that links the acceleration of the
scale factors of the subdomains to the global one: 
\begin{equation}
\frac{\ddot{a}_{\CD}}{a_{\CD}}=\sum_{\ell}\lambda_{\ell}\frac{\ddot{a}_{\ell}\left(t\right)}{a_{\ell}\left(t\right)}+\sum_{\ell\neq m}\lambda_{\ell}\lambda_{m}\left(H_{\ell}-H_{m}\right)^{2}\;.
\label{eq:acc-split}
\end{equation}
As an immediate consequence one can see that even when the subregions
decelerate, the second term of (\ref{eq:acc-split}) may counterbalance
the first one to lead to global accelerated expansion. We will examine
this property, which appears generically in averaged models due to correlations of local expansion rates,
in Appendix \ref{sub:A-first-example}.

In the following we will be mainly considering the case where we have
only two types of subregions $\CF_{\ell}$: over--
and underdense regions. We will define them at some initial time and
call $\CM:=\CF_{1}$ the part of the hypersurfaces that consist of all
elementary regions $\CM^{\left(\alpha\right)}:=\CF_{1}^{\left(\alpha\right)}$
with an initial overdensity and $\CE:=\CF_{2}$ the complementary
part $\CE:=\CD\cap\CM$, i.e. the one with the initially underdense
regions. The formulae (\ref{eq:split}) and (\ref{eq:Q-split}) then
simplify to give
\begin{equation}
H_{\CD}=\lambda_{\CM}\, H_{\CM}+\left(1-\lambda_{\CM}\right)H_{\CE}\;,
\label{eq:H-zerlegt}
\end{equation}
with the same expression valid also for $\average{\varrho}$ and $\average{\CR}$,
and
\begin{eqnarray}
\CQ_{\CD} & = & \lambda_{\CM}\CQ_{\CM}+\left(1-\lambda_{\CM}\right)\CQ_{\CE}
\label{eq:partQD}\\
 &  & +6\lambda_{\CM}\left(1-\lambda_{\CM}\right)\left(H_{\CM}-H_{\CE}\right)^{2}\;.\nonumber 
\end{eqnarray}
In both cases we used $\sum_{\ell}\lambda_{\ell}=1$, which results
from the fact that the $\CF_{\ell}$ are disjoint, and defined $\lambda_{\CM}:=\left|\CM\right|/\left|\CD\right|$
and $\lambda_{\CE}:=\left|\CE\right|/\left|\CD\right|$. This
general separation formula allows us to illustrate the simplifications
applied by Wiltshire in his two--scale model \cite{wiltshire:clocks}.
As he assumes the overdense $\CM$--regions to have no curvature,
by Eq. (\ref{eq:Integrability-curv}) the only possible nonzero $\CQ_{\CM}$--term
would have a strongly decaying $a_{\CM}^{-6}$--behavior and may
therefore equally well be set to zero. For the underdense $\CE$--regions
the situation is similar as he assumes them to have the Friedmann--like
$a_{\CE}^{-2}$ constant curvature term. This again makes the left--hand 
side of Eq. (\ref{eq:Integrability-curv}) vanish and leads to
the choice of $\CQ_{\CE}=0$. This reduces the general formula (\ref{eq:partQD})
to its third term on the right--hand side, which corresponds to his
Eq. (31) in \cite{wiltshire:clocks}, that he derived for the case
of vanishing shear. This is consistent, as we remarked below (\ref{eq:Q-split}),
because the shear term of $\CQ_{\CD}$ is completely contained in $\CQ_{\CM}$
and $\CQ_{\CE}$. Considering the overall formalism his choice is conservative
in the sense that one would indeed expect the biggest backreaction
effect to stem from the variance of the expansion rates between the
$\CM$-- and $\CE$--regions and not from those on $\CM$ and $\CE$
themselves, as they are chosen to have similar properties and therefore
similar expansion rates. On the other hand, setting the local backreaction
to zero restricts the new interesting feature of the averaging formalism,
i.e. the coupling of backreaction to curvature, to the $\CD$--regions
only. Furthermore, as we argue in Subsection~\ref{sub:numerical-example},
it may be eligible to allow for shear on the $\CM$--regions, which
may (and will) imply a negative $\CQ_{\CM}$ (i.e., a Dark Matter behavior over $\CM$). 
We will therefore not restrict ourselves
to the assumptions of constant curvature, but use other well--motivated
conditions in Section~\ref{sec:A-simple-scaling}.

\subsection{Consistent split of the dynamical equations}

With the knowledge of the relation between the quantities on the subregions
with those on the global domain $\CD$, one may now ask, how the separation
affects the evolution equations for those quantities. Therefore, we
insert the expressions (\ref{eq:split}) for $H_{\CD}$, $\average{\varrho}$
and $\average{\CR}$ and (\ref{eq:Q-split}) for $\CQ_{\CD}$ into (\ref{eq:Raychaudhuri-Mittel}),
(\ref{eq:Hamilton-Mittel}) and (\ref{eq:Integrability-curv}). A
straightforward calculation shows that the equations take the following
form: 
\begin{eqnarray}
0 & = & \sum_{\ell}\lambda_{\ell}\left[8\pi G\left\langle \varrho\right\rangle _{\ell}-\frac{{\CQ}_{\ell}+\left\langle \CR\right\rangle _{\ell}}{2}+\Lambda-3H_{\ell}^{2}\right]
\label{eq:Ham-Allgemein-Separiert}\\
0 & = & \sum_{\ell}\lambda_{\ell}\left[-4\pi G\left\langle \varrho\right\rangle _{\ell}+{\CQ}_{\ell}+\Lambda-3\frac{\ddot{a}_{\ell}\left(t\right)}{a_{\ell}\left(t\right)}\right]
\label{eq:Ray-Allgemein-Separiert}\\
0 & = & \sum_{\ell}\lambda_{\ell}\left[a_{\ell}^{-2}\partial_{t}\left(a_{\ell}^{2}\left\langle \CR\right\rangle _{\ell}\right)+a_{\ell}^{-6}\partial_{t}\left(a_{\ell}^{6}\CQ_{\ell}\right)\right]\;,
\label{eq:Int-Cond-Allgemein-Separiert}
\end{eqnarray}
i.e. they may be split into a sum in which the equations on the subregions
take the same form as those of the global region (\ref{eq:Raychaudhuri-Mittel})--(\ref{eq:Konservation-Mittel})
and their contribution is weighted with the volume fraction of the
respective subregion. As Equations (\ref{eq:Raychaudhuri-Mittel})--(\ref{eq:Konservation-Mittel})
hold for an arbitrary domain $\CD$ and therefore also for the subregions
$\CF_{\ell}$, Equations (\ref{eq:Ham-Allgemein-Separiert})--(\ref{eq:Int-Cond-Allgemein-Separiert})
show that the separation advocated in the previous section is also
consistent at the level of the evolution equations. This is not surprising
as the separation procedure is straightforward and the equations are
supposed to hold on any domain, but was, especially in view of the
nonlinear form of (\ref{eq:Q-split}), not entirely clear when just
looking at the formulas. The consistent split assures that, if we have
found a solution for the quantities on the subregions $\CF_{\ell}$
and use the relations of the previous section to calculate those on
the global domain, then we will automatically obtain a solution of the
averaged Equations (\ref{eq:Raychaudhuri-Mittel})--(\ref{eq:Konservation-Mittel})
on this global domain.

\section{Backreaction scenario based on extrapolating the leading perturbative mode
\label{sec:A-simple-scaling}}

After setting up the equations that link different subregions of a
certain partitioning of space to the whole in the preceding section,
we will now have a closer look at the simplest partitioning. As already
mentioned at the end of \ref{sub:Separation-formulae}, we are interested
in a subdivision into over-- and underdense regions. These will be labeled
by $\CM$ and $\CE$, respectively, and the classification is made at some
initial time and then kept fixed. As the averaged equations are derived
for a certain region $\CF_{\ell}$, that is assumed to keep its identity,
an exchange of elementary space entities $\CF_{\ell}^{\left(\alpha\right)}$
between the two classes of regions $\CM$ and $\CE$ would spoil the
applicability of (\ref{eq:Raychaudhuri-Mittel})--(\ref{eq:Konservation-Mittel})
on $\CM$ and $\CE$, and therefore would lead to some complicated modified
form which would introduce even more unknown parameters. This is the
reason for keeping the identification of $\CF_{\ell}$--regions,
even if this will introduce some problems with the clearcut interpretation
of $\CM$ and $\CE$, for initially overdense regions may become
underdense in the course of evolution. A refined description that
helps to evade this problem will be mentioned in the discussion in Sec.~\ref{sec:Discussion}.

The motivation of distinguishing between the $\CM$-- and $\CE$--classes of
regions is to get an explicit handle on structure formation. Implicitly
it is already present in the averaging formalism mainly via the $\left(\average{\theta^{2}}-\average{\theta}^{2}\right)$--term
in the kinematical backreaction defined by (\ref{eq:Def-Q}). As structure
formation implies a different evolution of the local expansion rate
at different places in the Universe, this variance is expected to
grow. The distinction now offers the possibility to examine the development
of parts that are in the focus of experimental
interest, namely "voids" (see e.g. \cite{voidfinder} and references therein)
and clusters of galaxies. Moreover, it allows us
to trace the process of structure formation by the simplest parameter
characterizing its history, i.e. the volume fraction of the overdense
regions. The current section presents the outcome of this investigation
for the special case of exact scaling solutions.

\subsection{Exact scaling solutions of the averaged equations}

As we have seen in Section~\ref{sub:The-average-Einstein}, the averaged
equations do not form a closed system. One closure strategy is to impose
a specific equation of state. This is similar to closing the Friedmannian equations, but here the
equation of state is dynamical. As done in quintessence models for the $X$--matter content inserted
as source into the standard Friedmann equations, the simplest assumption is
to choose a constant equation of state parameter $w$. This means
that we have a single scaling law for the corresponding matter component.
If we proceed in a similar way for the backreaction and curvature
terms in the averaged equations, we may choose (following \cite{morphon}):
\begin{equation}
\CQ_{\CF}=\CQ_{\CF_{i}}a_{\CF}^{n}\;;\;\langle\CR\rangle_{{\cal \CF}}=\CR_{\CF_{i}}a_{\CF}^{p}\;,
\label{eq:Skalierung}
\end{equation}
where $\CF$ stands for $\CM$ or $\CE$ and where $\CF_{i}$ indicates
the initial state of the domain $\CF$. By the integrability condition
(\ref{eq:Integrability-curv}), $\CQ_{\CF}$ and $\averageF{\CR}$ are
related. Inserting the above ansatz leads to two types of solutions:
first, there is a solution with $n=-6$ and $p=-2$. This one is
not very interesting as this is the only case where the generic coupling
of backreaction and curvature is absent. This case would correspond to fluctuation histories that are decoupled from a constant--curvature (Friedmann--like)
behavior. Second, (\ref{eq:Integrability-curv})
will be satisfied if $n=p$. This is the case we are going to study
in the following. It implies that $\CQ_{\CF}$ and $\averageF{\CR}$
are related by a constant $r^{\CF}$ 
\begin{equation}
\CQ_{\CF}=r^{\CF}\langle\CR\rangle_{{\cal \CF}}=r^{\CF}\CR_{\CF_{i}}a_{\CF}^{n}\;,
\label{eq:SkalierungN}
\end{equation}
which is determined by (\ref{eq:Integrability-curv}) to be 
\begin{equation}
r^{\CF}=-\frac{n+2}{n+6}\;.
\label{eq:Prop-Index-Skalierung}
\end{equation}
The effective equation of state for the $X$--matter (backreaction
and curvature) component, defined in (\ref{eq:eq-of-state}), is for
scaling solutions simply 
\begin{equation}
w_{\Lambda,\rm{eff}}^{\CF}=-\frac{1}{3}\left(n+3\right)\;.
\label{eq:Eq-state-scaling}
\end{equation}
For the explicit solution of the equations on $\CM$ and $\CE$ in
this paragraph we will further specify $n$ to $-1$. The motivation
for this choice is, that Li and Schwarz \cite{gaugeinv} found, in
a calculation using second--order perturbation theory, that the backreaction
term may be expressed as a Laurent series with a leading term going
like $a_{\CF}^{-1}$. We will rederive this behavior, under similar
conditions as in their case, in Section~\ref{sec:Power-series}. There
we also argue that the leading term of the scalar curvature, going
like $a_{\CF}^{-2}$ is expected to vanish for plausible initial conditions.
We will show that extrapolating the $a_{\CF}^{-1}$--behavior up to
today runs into trouble and that higher terms in the series are expected
to make an important contribution at the epoch of structure formation.
Nevertheless, the $a_{\CF}^{-1}$ scaling solution provides a first
and simple example of why one would expect that backreaction through
structure formation might cause accelerated expansion \cite{seikel:acc}.

We may argue that we did not have to rely on assumptions on the effective equation of state, if we exploited the knowledge of structure formation further.
Such an attempt runs into problems as we do not have an inhomogeneous relativistic metric model at our disposal that would cover the details of  nonlinear structure formation. On large scales we probably understand what is happening, but we are already using this information by the known $a_{\CF}^{-1}$ scaling behavior on these scales. In the late Universe with density contrasts
of order $1$, the closest to a realistic model at present is the one by R\"as\"anen \cite{rasanen:peakmodel}. But also this has to rely on some unrealistic assumptions.
We therefore try to derive the equation of state from the N--body simulation of structure formation in our second approach in Section~\ref{sec:Modelling-structure-formation}, where we use
the knowledge on nonlinear structure formation encoded in the volume fraction of the simulation.

\subsection{Free parameters and constraints\label{sub:Free-parameters}}

In order to derive the evolution of the global domain $\CD$ under
the assumption of the $a_{\CF}^{-1}$--scaling behavior for $\CM$
and $\CE$, we have to solve Equation (\ref{eq:Hamilton-Mittel})
(the other independent Equation (\ref{eq:Integrability-curv}) is
already satisfied by the scaling ansatz) which simplifies to
\begin{equation}
H_{\CD_{0}}^{2}\left[\Omega_{m}^{\CF_{0}}\frac{a_{\CF_{0}}^{3}}{a_{\CF}^{3}}+\Omega_{\CR \CQ}^{\CF_{0}}\left(\frac{a_{\CF_{0}}}{a_{\CF}}\right)^{-n}\right]=\left(\frac{\dot{a}_{\CF}}{a_{\CF}}\right)^{2}
\label{eq:Int-M-Ham}
\end{equation}
 for $\CF$ out of $\CM$ and $\CE$, where we have defined $\Omega_{\CR \CQ}^{\CF_{0}}:=\left(\Omega_{\CR}^{\CF_{0}}+\Omega_{\CQ}^{\CF_{0}}\right)$.
Instead of using $a_{\CM_{0}}$ and $a_{\CE_{0}}$ as free parameters
we may equally well work with $\lambda_{\CM_{0}}$ and $a_{\CD_{0}}$.
To specify these and the other free parameters, namely $H_{\CD_{0}}^{2}$,
$\Omega_{m}^{\CF_{0}}$, $\lambda_{\CM_{0}}$, $a_{\CD_{0}}$ and
$\Omega_{\CR \CQ}^{\CF_{0}}$ -- with the index $0$ denoting the value
of today -- we have to make some further assumptions. One is that
the density fluctuations in the Early Universe around $z=1000$, where
we choose our initial time $t_{i}$, were Gaussian and very small
as indicated by the results of the CMB experiments. For our parameters
this means that $\lambda_{\CM_{i}}\approx0.5$ as Gaussian fluctuations
imply that there were as many over-- as underdense regions and we
set $a_{\CM_{i}}=a_{\CE_{i}}=\sqrt[3]{1/2}$. The assumption
of being close to homogeneity implies $\langle\varrho\rangle_{\CD_{i}}\approx\langle\varrho\rangle_{\CM_{i}}\approx\langle\varrho\rangle_{\CE_{i}}$.
It then follows that
\begin{equation}
\Omega_{m}^{\CF_{0}}=\frac{8\pi G}{3H_{\CD_{0}}^{2}}\frac{a_{\CF_{i}}^{3}}{a_{\CF_{0}}^{3}}\langle\varrho\rangle_{\CF_{i}}\approx\lambda_{\CM_{i}}\frac{a_{\CD_{0}}^{3}}{a_{\CF_{0}}^{3}}\Omega_{m}^{\CD_{0}}\;,
\end{equation}
which allows us to simplify Equation (\ref{eq:Int-M-Ham}) by replacing
$\Omega_{m}^{\CM_{0}}$ and $\Omega_{m}^{\CE_{0}}$ with $\Omega_{m}^{\CD_{0}}$.
To reduce the set of unknown parameters even further, we may use (\ref{eq:Int-M-Ham})
today which gives
\begin{equation}
\frac{\lambda_{\CM_{i}}}{\lambda_{\CM_{0}}}\Omega_{m}^{\CD_{0}}+\Omega_{\CR \CQ}^{\CM_{0}}=\frac{H_{\CM_{0}}^{2}}{H_{\CD_{0}}^{2}}\;;\;\frac{\lambda_{\CM_{i}}}{1-\lambda_{\CM_{0}}}\Omega_{m}^{\CD_{0}}+\Omega_{\CR \CQ}^{\CE_{0}}=\frac{H_{\CE_{0}}^{2}}{H_{\CD_{0}}^{2}}\,,
\end{equation}
and may be used together with (\ref{eq:H-zerlegt}) to eliminate
$\Omega_{\CR \CQ}^{\CM_{0}}$. Another consistency condition finally
allows us to get rid of the parameter $\Omega_{\CR \CQ}^{\CE_{0}}$
as well, enabling us to fix our model without having to know anything
about the values of the backreaction or curvature terms.
The argument is as follows: consider the integral of Equation
(\ref{eq:Int-M-Ham}) 
\begin{equation}
H_{\CD_{0}}\intop_{0}^{t}dt^{\prime}=\intop_{\sqrt[3]{1/2}}^{a_{\CF}}\left[\frac{\Omega_{m}^{\CD_{0}}}{2}\frac{a_{\CD_{0}}^{3}}{a_{\CF}^{\prime}}+\Omega_{\CR \CQ}^{\CF_{0}}a_{\CF_{0}}a_{\CF}^{\prime}\right]^{-\frac{1}{2}}da_{\CF}^{\prime}\;,
\end{equation}
which gives two functions $t_{\CM}\left(a_{\CM}\right)$ and $t_{\CE}\left(a_{\CE}\right)$
depending on the parameters in the integral. This may be stated as
$t_{\CF}\left(a_{\CF}\right)=t_{\CF}\left(a_{\CF};\Omega_{\CR \CQ}^{\CF_{0}},H_{\CD_{0}},\Omega_{m}^{\CD_{0}},a_{\CD_{0}},\lambda_{\CM_{0}}\right)$.
As there is only one time even in the separated equations, which is
the time of the foliation, $t_{\CM}$ and $t_{\CE}$ have to be equal. (Note that here we do not introduce a phenomenological lapse between times of different regions as done by Wiltshire, since this would require another
(very involved) implementation of a multiscale foliation; we so must imply that Wiltshire's effect is negligible, but we do not claim that this is indeed the case.) 
Using this condition today, i.e. demanding that $t_{\CM}\left(a_{\CM_{0}}\right)=t_{\CE}\left(a_{\CE_{0}}\right)$
gives another relation linking $\Omega_{\CR \CQ}^{\CM_{0}}$ and $\Omega_{\CR \CQ}^{\CE_{0}}$
to the rest of the parameters. We may therefore eliminate $\Omega_{\CR \CQ}^{\CE_{0}}$
as well, which leaves us with the four free parameters $H_{\CD_{0}},\Omega_{m}^{\CD_{0}},a_{\CD_{0}}$
and $\lambda_{\CM_{0}}$. $a_{\CD_{0}}$ and $H_{\CD_{0}}$ only determine
the scaling of the axis so that the parameters responsible for the
shape of the curve of time evolution of $a_{\CM}\left(t\right)$ and
$a_{\CE}\left(t\right)$ are $\Omega_{m}^{\CD_{0}}$ and $\lambda_{\CM_{0}}$.

\subsection{A specific numerical example\label{sub:numerical-example}}
\begin{figure*}
\includegraphics[width=0.47\textwidth]{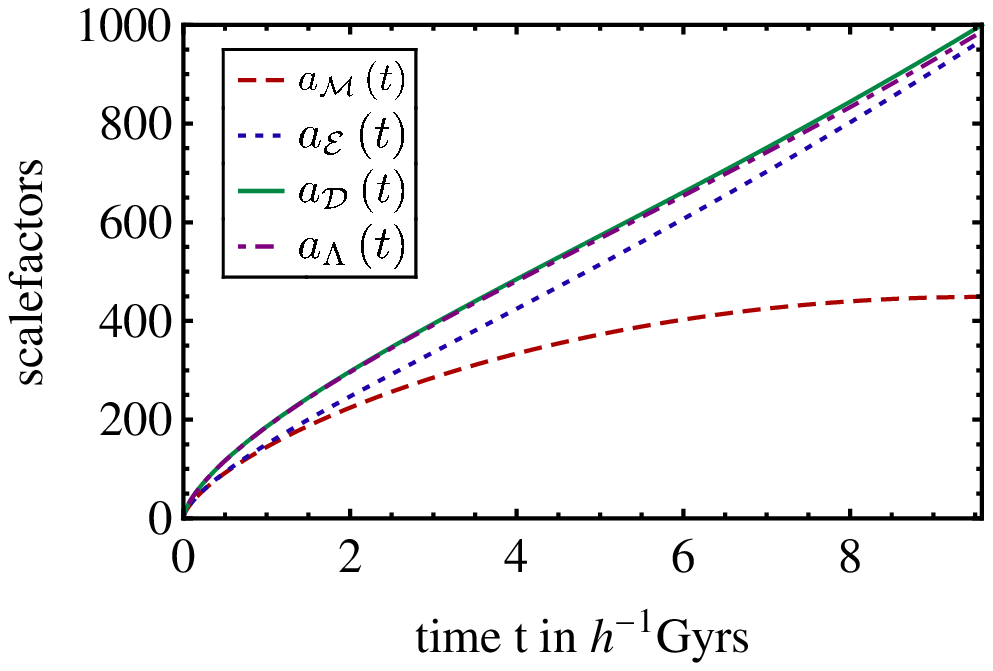}%
\hspace{0.05\textwidth}%
\includegraphics[width=0.47\textwidth]{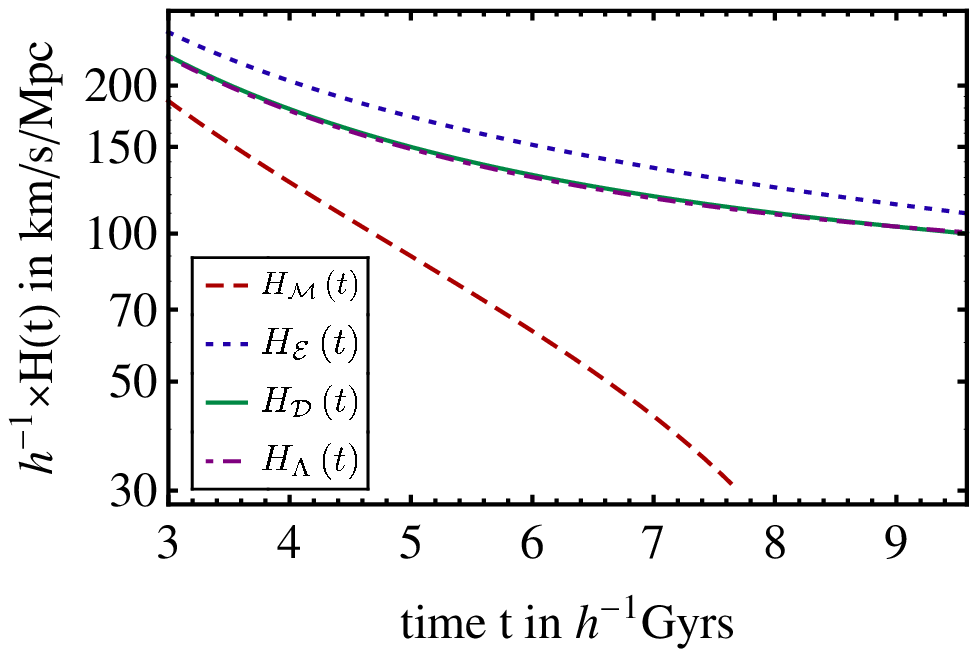}

\vspace{0.03\textwidth}%
\includegraphics[width=0.47\textwidth]{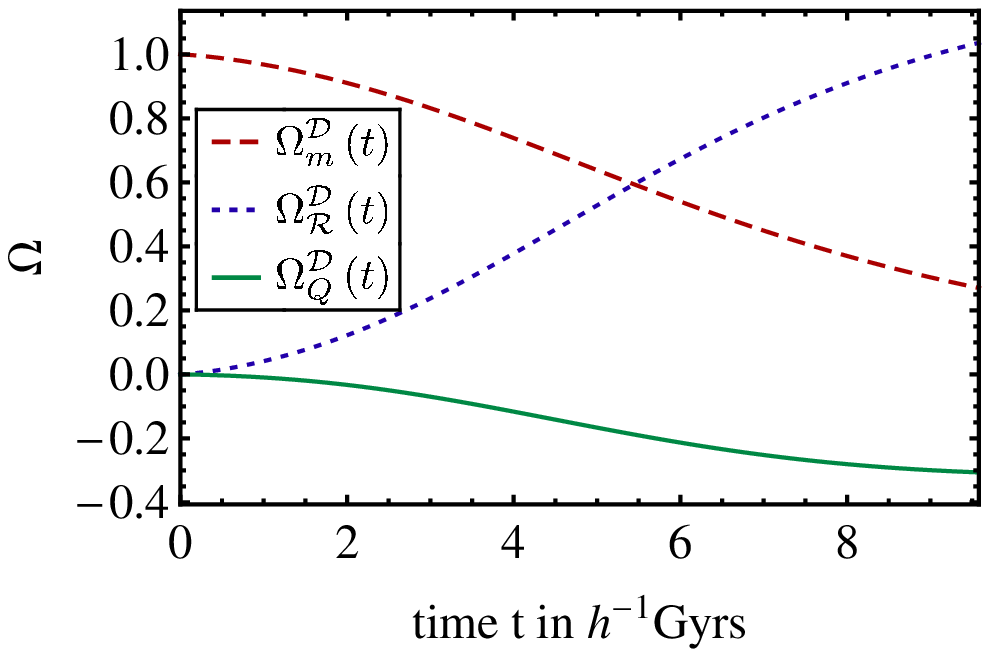}%
\hspace{0.05\textwidth}%
\includegraphics[width=0.47\textwidth]{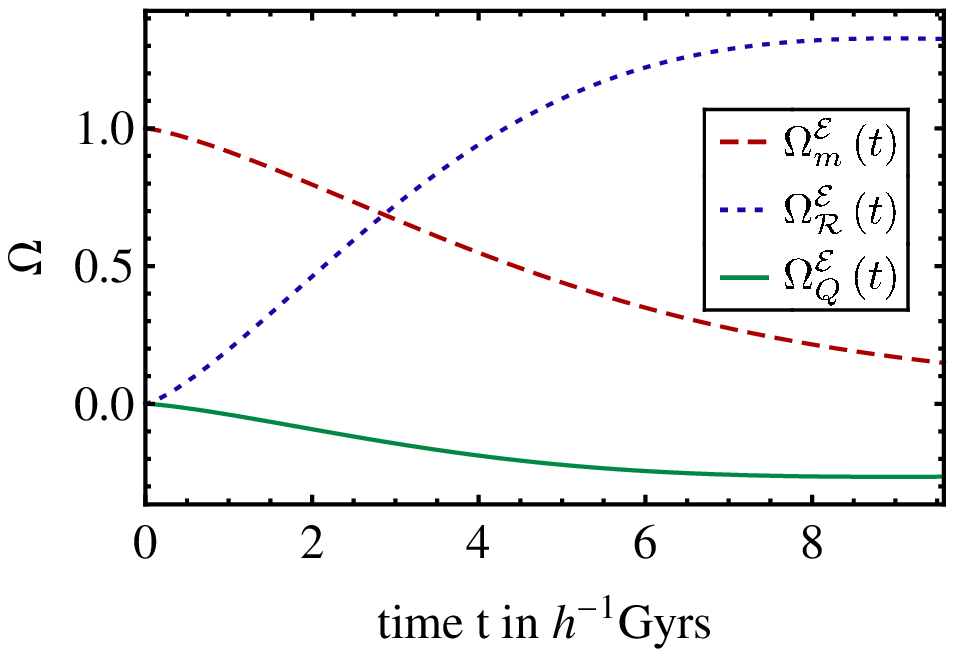}

\vspace{0.03\textwidth}%
\includegraphics[width=0.47\textwidth]{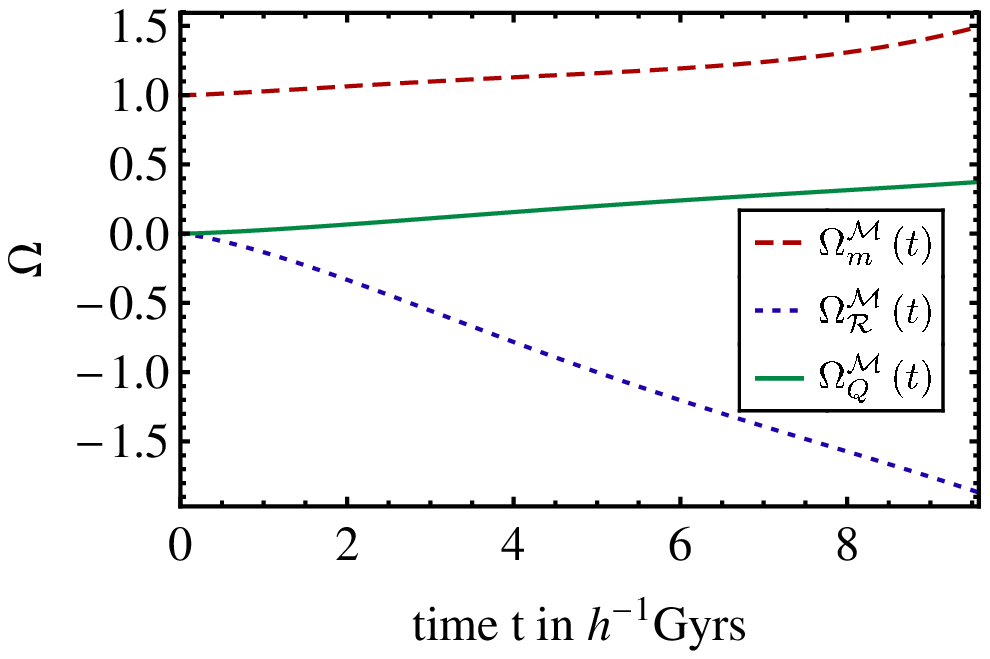}%
\hspace{0.05\textwidth}%
\includegraphics[width=0.47\textwidth]{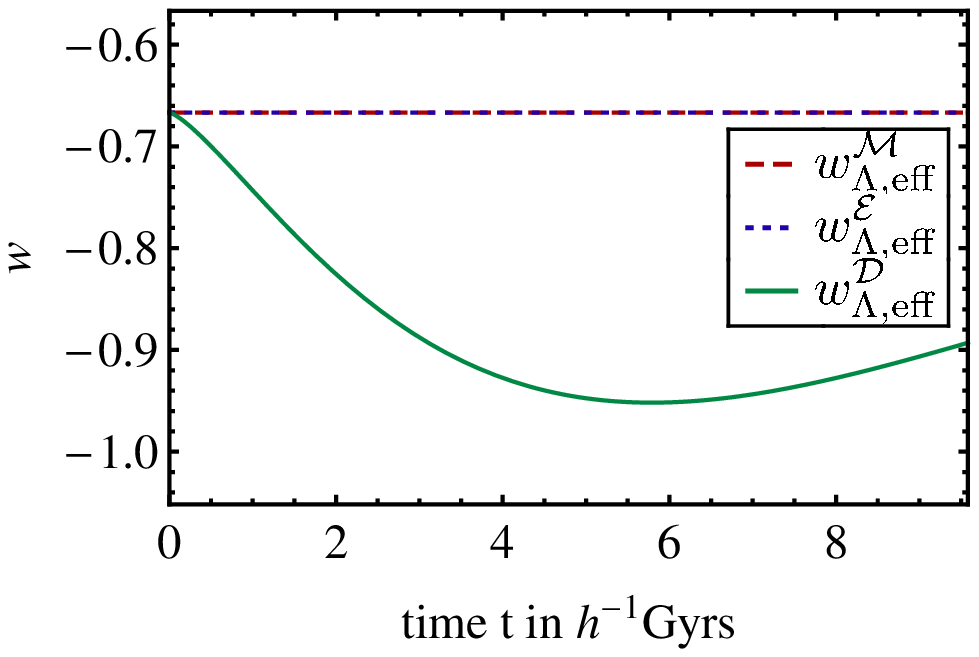}

\caption{Plots of some quantities derived from the single scaling model with
$\lambda_{\CM_{0}}=0.09$ and $\Omega_{m}^{\CD}=0.27$. Top
left: Comparison of the evolution of the different scale factors of
our model and the one in the $\Lambda$CDM--case. On the right the
corresponding Hubble rates. The one on $\CM$ goes to zero in the
end whereas $H_{\CD}$ approaches $H_{\CE}$ asymptotically. $H_{\Lambda}$
and $H_{\CD}$ are very close to each other. The middle row
shows the $\Omega$--parameters on $\CD$ and $\CE$.
Those for $\CM$ are left at the bottom. Finally, on the right
the behavior of the effective equation of state as defined in (\ref{eq:eq-of-state}).
The $a_{\mathcal{F}}^{-1}$--scaling implies that  $w_{\Lambda,\rm{eff}}^{\CM}$
and $w_{\Lambda,\rm{eff}}^{\CE}$ both have a value of $-\frac{2}{3}$.\label{fig:Graphen01}}

\end{figure*}

In order to get some quantitative statements we will specify the remaining
parameters to some plausible values. For $\Omega_{m}^{\CD_{0}}$ this
is the one indicated by the experiments that established the cosmological
concordance model. Even if it is not a priori clear that the domain
dependent $\Omega_{m}^{\CD_{0}}$ of the averaged equations should
have the same value as the $\Omega_{m}$ parameter whose value is
determined from the data under the assumption of a homogeneous and
isotropic universe model, this is the only thing we can do before the analysis
of observations in the context of the averaged model, begun in \cite{morphon:obs},
is completed. In this spirit we will use for $\lambda_{\CM_{0}}$
some reasonable value derived from a $N$--body simulation of large--scale structure
formation. These are in our context not the best sources of data either,
as they are Newtonian and therefore do not include scalar curvature
being a major player in the averaged equations. The reason why
we think that the value of the parameter $\lambda_{\CM_{0}}$ is nevertheless
derivable from them, is first, that they are designed in a way to
match best today's Universe and secondly, that, for the calculation
of the volume, even a substantial curvature deviation from a Euclidean space 
does not change the value very much.
This is because the calculation of volumes is performed with the metric
and its spatial derivatives that are linked to
curvature do not enter in the comoving foliation used here (see \cite{estim} for clarifying remarks on that point).
\begin{figure*}
\includegraphics[width=0.47\textwidth]{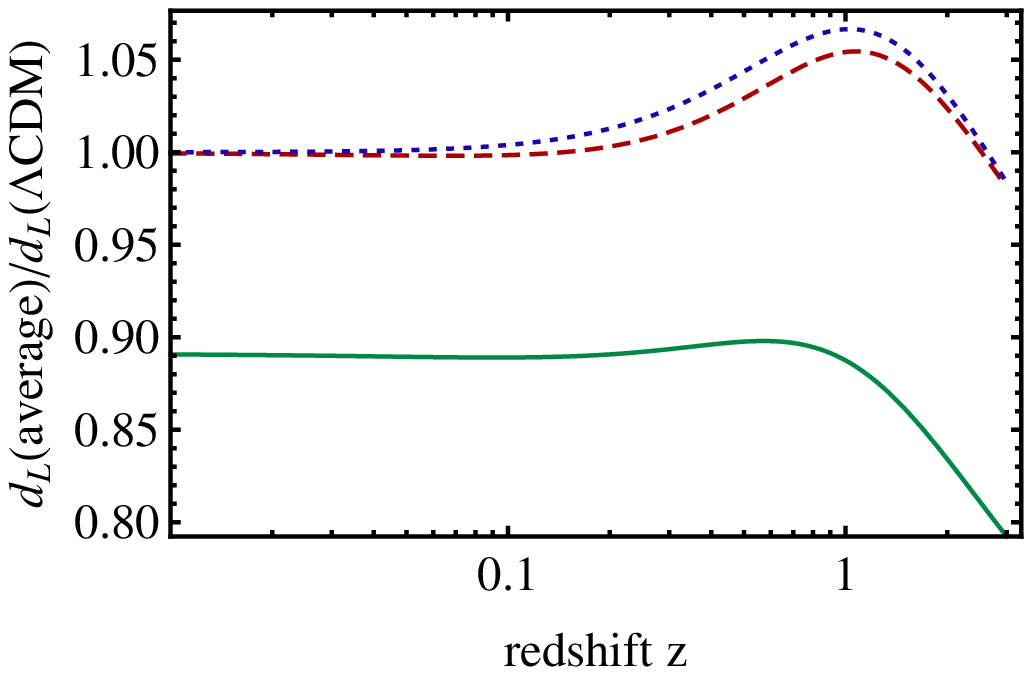}%
\hspace{0.05\textwidth}%
\includegraphics[width=0.47\textwidth]{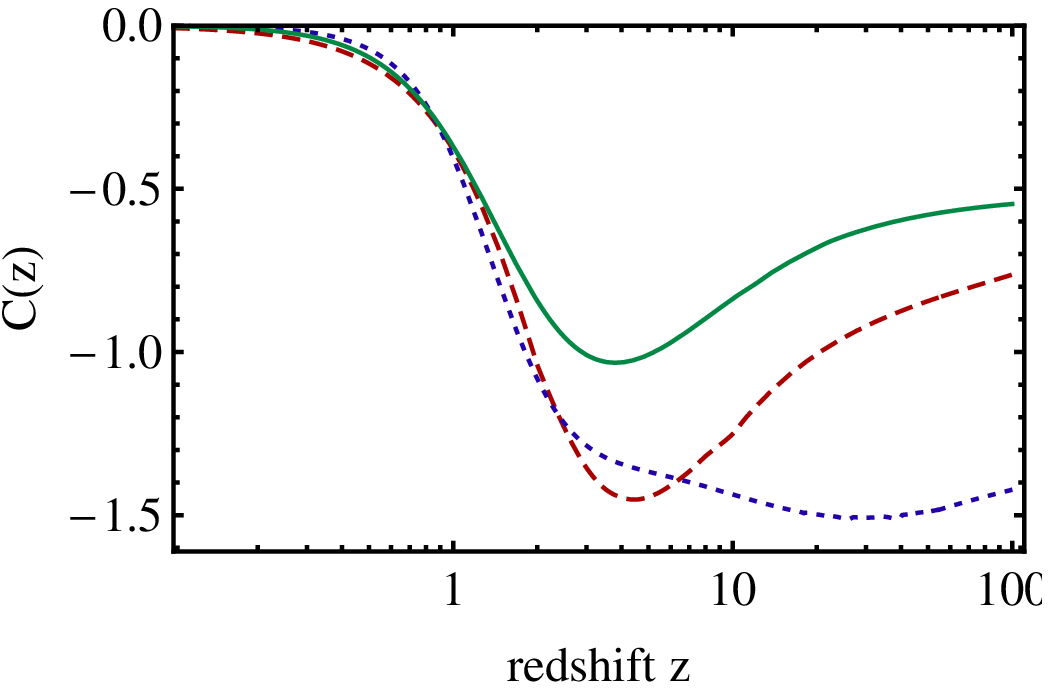}

\caption{Left: Comparison of the luminosity distances of our models (based on the template metric of \cite{morphon:obs}), with the one of a flat $\Lambda$CDM model with $h=0.7$ and $\Omega_{m}=0.27$. On top the model of Section~\ref{sec:Modelling-structure-formation}, where we force the volume scale factor $a_{\CD}$ to follow the $\Lambda$CDM evolution. Despite this assumption, the changing curvature affects the luminosity distance. The luminosity distances in our models show a significant feature at a redshift of around $1$, when compared with the best fit $\Lambda$CDM model, which may be looked for in the SN data. The curve below is the model of the current section. For comparison we also included the luminosity distance of the best fit model of \cite{morphon:obs}. Because of a different Hubble rate of $h=0.7854$ it lies below the others from the beginning. This model does not significantly show the distinct feature of the other two models around a redshift of $1$, due to the assumption of a single--scale cosmology.\\Right: Values of Clarkson's $C$--function \cite{clarkson} for the best fit model of \cite{morphon:obs} (top), the model of Section~\ref{sec:Modelling-structure-formation} (middle), and the model of this section (bottom). Recall that for every Friedmann model $C(z)$ vanishes exactly on all scales and for all redshifts. For the inhomogeneous models shown in the plot, this function has a minimum which may serve as observational evidence for the effective cosmologies, as proposed in \cite{morphon:obs}. As our multiscale model shows, it is not even necessary to measure derivatives of distance, since the feature is already present in the distance. \label{fig:lumidist}}

\end{figure*}

We will explain further details of our analysis of the $N$--body data
in Appendix \ref{sub:A-simple-mesh}. The value we will be using in
this section has been obtained by dividing the analysis volume into
blocks with sidelength $5\rm{h^{-1}Mpc}$ and evaluating whether
they are over-- or underdense. This leads to $\lambda_{\CM_{0}}\approx0.09$.
Together with $\Omega_{m}^{\CD_{0}}\approx0.27$ and $a_{\CD_{0}}\approx1000$
we may numerically evaluate Equation (\ref{eq:Int-M-Ham}) which
results in the curves of Fig.~\ref{fig:Graphen01}. The value of $a_{\CD_{0}}\approx1000$
is used to identify {}``today'' in our plots. It can be interpreted in line with the standard model,
if the global volume is comparable, i.e. if perturbations of the metric remain small.
In any case, substantial changes are expected for the spatial derivatives of the metric, 
and consequently for the time--derivatives of the scale factor.

\subsubsection{Evolution of the model universe}

The top left graph of Fig.~\ref{fig:Graphen01} shows the evolution of the different scale factors
with cosmic time. For comparison it also contains a plot of the scale
factor $a_{\Lambda}\left(t\right)$ for a flat standard FLRW model
with a matter content of $\Omega_{m}=0.27$ and $73\%$ Dark Energy.
One can see that its evolution is very similar to the one of the volume
scale factor $a_{\CD}\left(t\right)$ of the global domain $\CD$
throughout the whole history. 

That this apparent match is also expected to hold true for observables is suggested by the analysis in Fig.~\ref{fig:lumidist}, where we compare the luminosity distances at a given redshift of the
average model and the standard $\Lambda$CDM model. To calculate the distances we used the effective template metric of \cite{morphon:obs}, which provides a prescription of how to incorporate
the changes in the average scalar curvature for the calculation of distances. We see that for the region where we have ample SN data, i.e. around a redshift of $1$, the difference between our model and the $\Lambda$CDM model is less than 5\%.

Considering the evolution of $a_{\CM}$ and $a_{\CE}$ we find that
they behave as one would expect from their nature. The overdense regions
$\CM$ begin their evolution as the underdense $\CE$--regions due
to their assumed similar matter content. Then their density increases
by gravitational instability and the expansion slows down. The underdense
regions which contain most of the voids continue their expansion.
This may also be seen from the Hubble rates in the plot on the right.
The expansion rate of the overdense regions drops down to zero rather
rapidly whereas the one of the underdense regions slows down much less.
The shrinking percentage of the overdense regions causes $H_{\CD}$
to bend up again implying acceleration. The difference to the FLRW
Hubble rate is not yet visible, but as the evolution proceeds the cosmological
constant of the FLRW model will force $H_{\Lambda}$ to rise in the
future whereas $H_{\CD}$ will join the evolution of $H_{\CE}$.

\subsubsection{Energy content}

The plot of the $\Omega$--parameters in Fig.~\ref{fig:Graphen01} is consistent with the conclusion drawn from Eq. (\ref{eq:Raychaudhuri-Mittel}) that $\Omega_{\CQ}^{\CD}$ has to be less than $-\frac{1}{4} \Omega_{m}^{\CD}$ to give accelerated expansion. In addition, the magnitude of $\Omega_{\CR}^{\CD}$ underlines the statement that curvature is an important player in our model. The combination of these two facts has an interesting consequence: if, even for a small but nonzero backreaction parameter, the average scalar curvature parameter is of order $1$, and therefore the actual matter parameter $\Omega_{m}^{\CD}$ is small, the model may explain Dark Energy without a large amount of backreaction. A small $\Omega_{\CQ}^{\CD}$ larger than $-\frac{1}{4} \Omega_{m}^{\CD}$ does, however, not explain an accelerated expansion, but this latter property may actually not be needed to explain observations. This remark is particularly interesting in view of the observation in Subsection~\ref{sub:baryons-only}.

That curvature is in fact more important
than suggested by the restricted constant curvature FLRW models has
been shown recently in \cite{estim}. The plots in the following row
clarify its origin. It stems from the underdense regions which are
supposed to contain most of the large voids that have developed negative
curvature in the course of evolution. The value for the matter parameter
of about $0.15$ makes clear that one may not think of $\CE$ being
composed of voids only. From its origin as the underdense \emph{half}
of the initial universe volume, it should be clear that it also has to contain
regions that have developed overdensities.

For the overdense regions the situation is different in that in the course
of their evolution, positive curvature emerges. This is what one would
expect due to the fact that they contain many gravitationally bound
systems like clusters. Another observation that coincides with the
intuitive picture we have, is the fact, that the backreaction contribution
is negative [note again the sign in the definition (\ref{eq:Def-Omega-Parameter})].
According to definition (\ref{eq:Def-Q}), negative backreaction means that
the region over which we average is shear dominated. In view of the
filamentary structure of the matter distribution of the Universe this
is also what we would expect.

When interpreting the $\Omega$--parameters of $\CM$ and $\CE$ one
should have in mind that they do not have to add up to unity but,
due to the definition of the $\Omega$'s using $H_{\CD}$ and not
$H_{\CM}$ and $H_{\CE}$, will add up to $H_{\CM}^{2}/H_{\CD}^{2}$
and $H_{\CE}^{2}/H_{\CD}^{2}$. This means for $\CM$ that the sum
at the end will be close to $0$, whereas the value on $\CE$ approaches
$1$.

Finally, the plot in the last row on the right shows the evolution
of the effective equation of state for the $X$--Matter component.
For $\CM$ and $\CE$ the value is, according to (\ref{eq:Eq-state-scaling}),
simply $-2/3$. For the global domain $\CD$ we have an evolution
from this limiting value to $-0.95$ at the maximum of structure formation
which then relaxes back to $\approx0.9$ today. It is interesting
to note that the value stays relatively constant throughout the period
of structure formation. %

\begin{figure}
\includegraphics[width=0.45\textwidth]{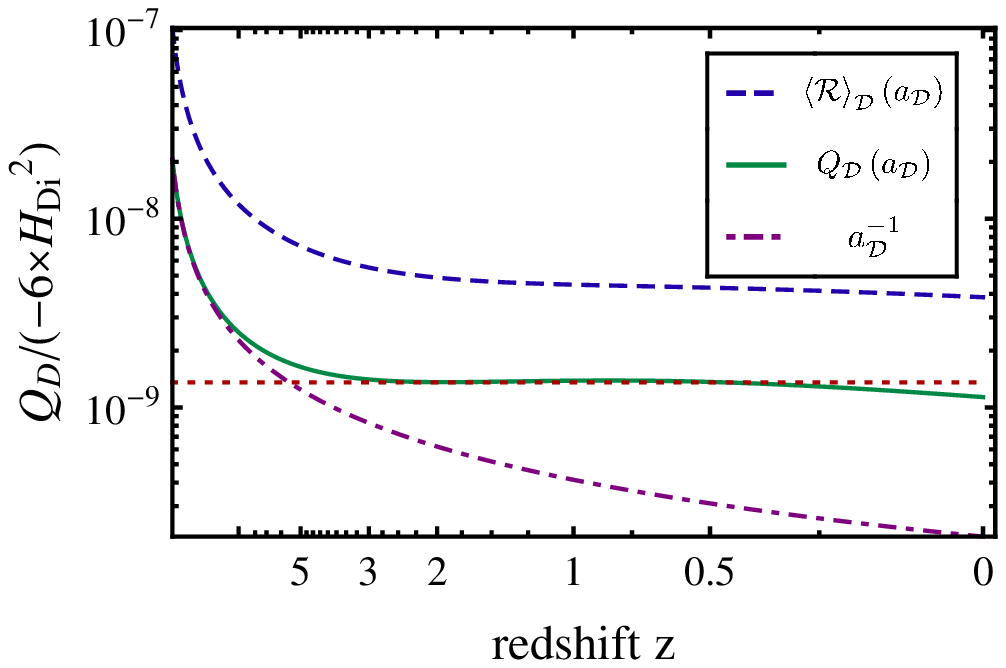}

\caption{Plot of the evolution of $\CQ_{\CD}$ and $\average{\CR}$ in terms of the global scale factor $a_{\CD}$. For comparison we added a line with a simple $a_{\CD}^{-1}$--scaling and one that is constant. $\CQ_{\CD}$ and $\average{\CR}$ are normalized by $-6H_{\CD_{\rm{i}}}^{2}$ so that the values at the initial time represent $\Omega_{\CQ}^{\CD_{\rm{i}}}$ and $\Omega_{\CR}^{\CD_{\rm{i}}}$. We appreciate that the backreaction terms feature an approximate cosmological constant behavior on the homogeneity scale despite the assumption of $a_{\CF}^{-1}$--scaling on the partitioned domains. Physically, this result can be attributed to the expansion variance between the subdomains and, hence, this latter is identified as the key effect to produce a global Dark Energy--like behavior of the backreaction terms.
\label{fig:Scaling-Q-R}}

\end{figure}

This observation, i.e. that the $a_{\CF}^{-1}$--scaling model on the partitioned domains naturally leads to an approximate cosmological constant behavior on the homogeneity scale, may also be seen from Fig.~\ref{fig:Scaling-Q-R}. There it is shown, how the backreaction and curvature terms deviate from their initial $a_{\CD}^{-1}$--scaling when structure formation sets in. They stay approximately constant for quite a long period of the evolution before the $\CE$--regions finally dominate and the backreaction and curvature terms begin to decay faster again.

If it were easily possible to interpret 
the quantities of the averaged model in terms of standard model ones, one might well imagine 
the value of the equation of state today of Fig.~\ref{fig:Graphen01} to lie within the $1-\sigma$ boundary derived by the WMAP team for a time--dependent dark energy equation of state \cite{wmap}.
We do not show the
corresponding plot however, because the comparison is not expected to
be very meaningful as the analysis uses the FLRW model as a prior.
We again have to ask the reader to await a more reliable
interpretation of the data in the context of the averaged model.

\subsubsection{Dark Matter\label{subsub:dark-matter}}

The comparison of the $\Omega$--evolution of the $\CM$-- and $\CE$--regions
in Fig.~\ref{fig:Graphen01} illustrates nicely the property of averaged
models that, depending on the domain under consideration, the effective
energy content may vary. Seen in terms of $X$--matter this means that
this component behaves in a more Dark Energy--like way when we consider
only the $\CE$--regions, since the backreaction is positive and drives
acceleration. For the $\CM$--regions, however, the backreaction term
is negative and slows down the evolution in the same manner as a Dark
Matter component would act \cite{buchert:new}. If backreaction is an important contribution
in today's Universe, this might imply that parts of the effect known
as Dark Matter could be due to a combination of the backreaction and
curvature term on $\CM$--regions. This would perhaps open the possibility to
reinterpret Dark Matter models in which the particles decay and the Dark Matter
responsible for the acoustic oscillations might be replaced through its
cosmological effect by a rising $X$--matter component.

It is to be expected that only parts of the effect attributed to Dark Matter is modeled by the effect of a negative backreaction, since there
are many independent lines of evidence for the existence of Dark Matter. While part of the Dark Matter problem related to the expansion history might well be attributed to backreaction, this would only alter the Late Universe constraints. The relative abundances of Dark Matter to baryonic matter in the Early Universe, determined by the CMB, are not expected
to change if we assume a near--homogeneous initial state of the Universe. All CMB constraints that stem from a combination of primordial information and constraints derived from observations of the Late Universe, however, will be subject to changes in course of a reinterpretation of observational data.
To show what conditions the CMB data really impose, and that will have to be satisfied also by a backreaction model, we refer to a recent paper \cite{cmbobs} that analyses the CMB in a model independent way.

If we choose for the $X$--matter component the description of \cite{morphon}
in terms of an effective scalar field, the {}``morphon'', we would
have two effective potentials depending on whether
we are in an over-- or underdense region. They are of the form 
\begin{eqnarray}
u_{\CM}\left(s_{\CM}\right) & = & -\frac{4}{\sqrt{5}}H_{\CM}^{2}\Omega_{m}^{\CM}\left(-\gamma_{\CR m}^{\CM}\right)^{\frac{3}{2}}\sin^{-1}\left(s_{\CM}\right)\;,\\
u_{\CE}\left(s_{\CE}\right) & = & \frac{4}{\sqrt{5}}H_{\CE}^{2}\Omega_{m}^{\CE}\left(\gamma_{\CR m}^{\CE}\right)^{\frac{3}{2}}\sinh^{-1}\left(s_{\CE}\right)\;,
\end{eqnarray}
 where $\gamma_{\CR m}^{\CF_{i}}:=\Omega_{\CR}^{\CF_{i}}/\Omega_{m}^{\CF_{i}}$,
$s_{\CF}\left(t\right):=\sqrt{8\pi G}\Phi_{\CF}\left(t\right)$ and
$u_{\CF}\left(s\right):=8\pi G\, U_{\CF}\left(\Phi\right)$. These
potentials are simply the specialization of the general result (37)
of \cite{morphon} to the $a_{\CF}^{-1}$--scaling and the different
negative sign for $\CQ_{\CM}$. The $\sin^{-1}$--behavior reflects
the fact that formally the $\CM$--regions recollapse. In reality
the description as a comoving perfect fluid will, however, break down
before so that only the rising branch is expected to be physical.
The values of the parameters on $\CM$ (more precisely the fact that $\average{\CR}$ is positive and $n=-1$, see \cite{morphon} for details) imply, that we have an effective
phantom field for the overdense regions. In spite of the usual interpretation in terms of Dark Energy, the different signs lead to a positive effective pressure and therefore to a decelerating component, which hence rather acts in a matterlike way.
The form of the potential
for the parameters of the model of Fig.~\ref{fig:Graphen01} has been
plotted in \cite{wiegand}. The approach to describe the effect of
inhomogeneities by an effective scalar field allows us to connect the
results obtained by the quintessence and scalar field community to
the backreaction formalism and to reinterpret their fields and potentials
in terms of physical quantities, i.e. the parameters of the averaged
model. This is potentially interesting as $\sinh^{\beta}$--potentials
have been shown to be able to behave like a Dark Energy component, e.g. in \cite{matos} or \cite{sahni} [where their $\left(\cosh-1\right)^{p}$ potential is just a $\sinh^{\beta}$ one by $\cosh\left(2x\right)-1=2\sinh^{2}\left(x\right)$]. For more detailed information to possible quintessence potentials, see \cite{DE:review} and references therein.

Note that scalar fields have also been used to model Dark Matter halos \cite{matos:halos} and that other potentials for a scalar field have been advocated to implement the cosmological evolution of such scalar field Dark Matter \cite{matos:dm,matos:dm09}. Consequently, it has been proposed to unify these Dark Matter and Dark Energy scalar fields into a single description e.g. in \cite{arbey} and \cite{matos:unify}. Questions related to unification have been also addressed in a recent work on employing the Chaplygin gas as effective equation of state in the present context \cite{chaplygin}.
Note also that the common strangeness of scalar field models, i.e. requiring particles with masses of order of $10^{-26}\rm{eV}$ is not present in the morphon picture as there the notion of a {}``particle'' is only an effective one.

\begin{figure*}
\includegraphics[width=0.47\textwidth]{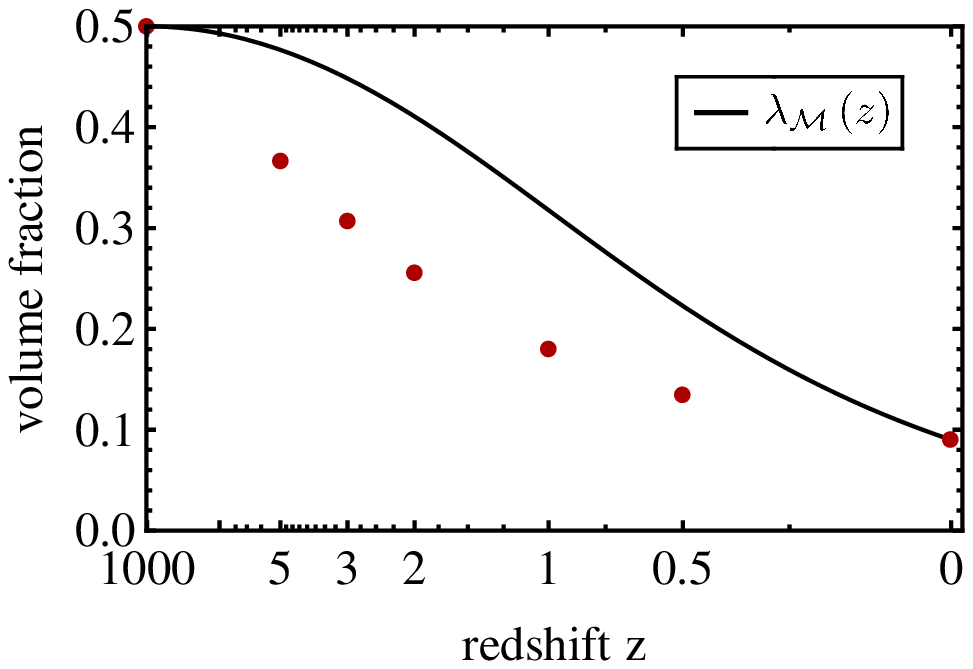}%
\hspace{0.05\textwidth}%
\includegraphics[width=0.47\textwidth]{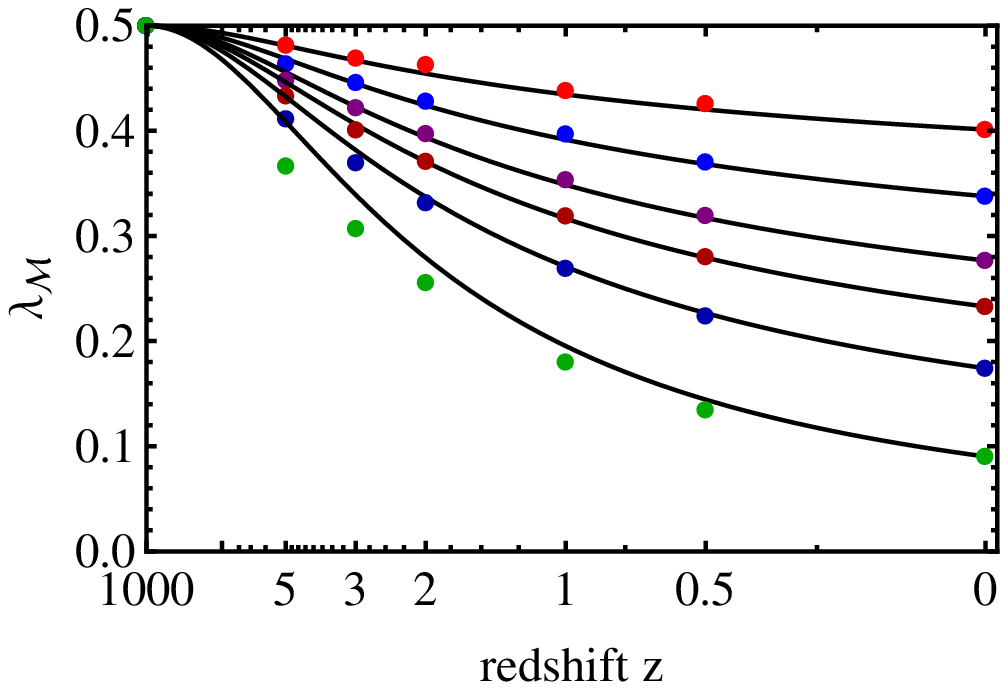}

\vspace{0.05\textwidth}%
\includegraphics[width=0.47\textwidth]{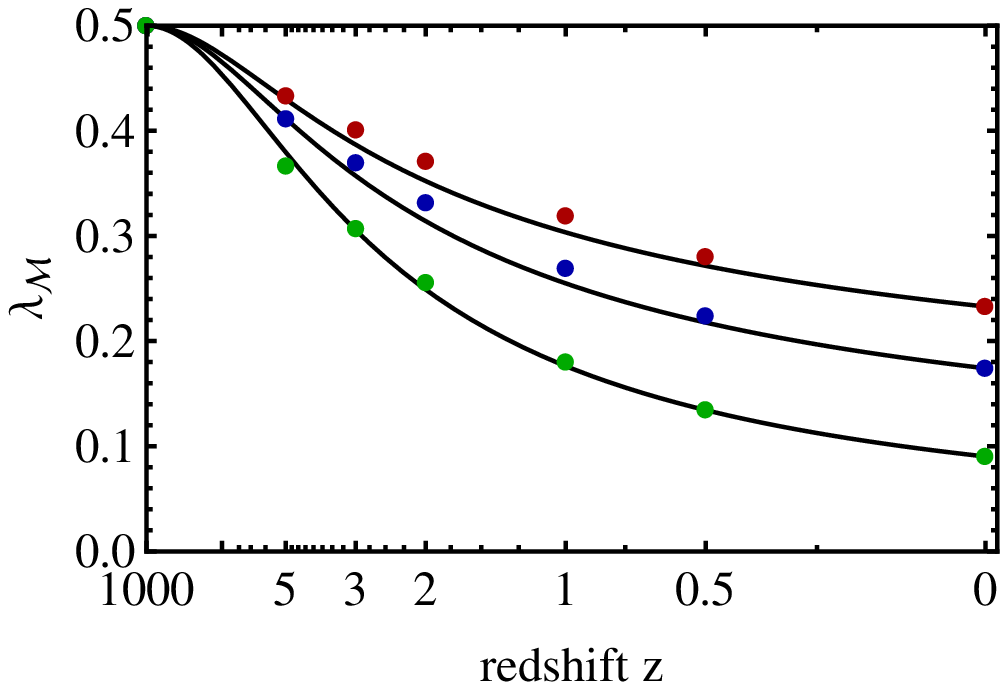}%
\hspace{0.05\textwidth}%
\includegraphics[width=0.47\textwidth]{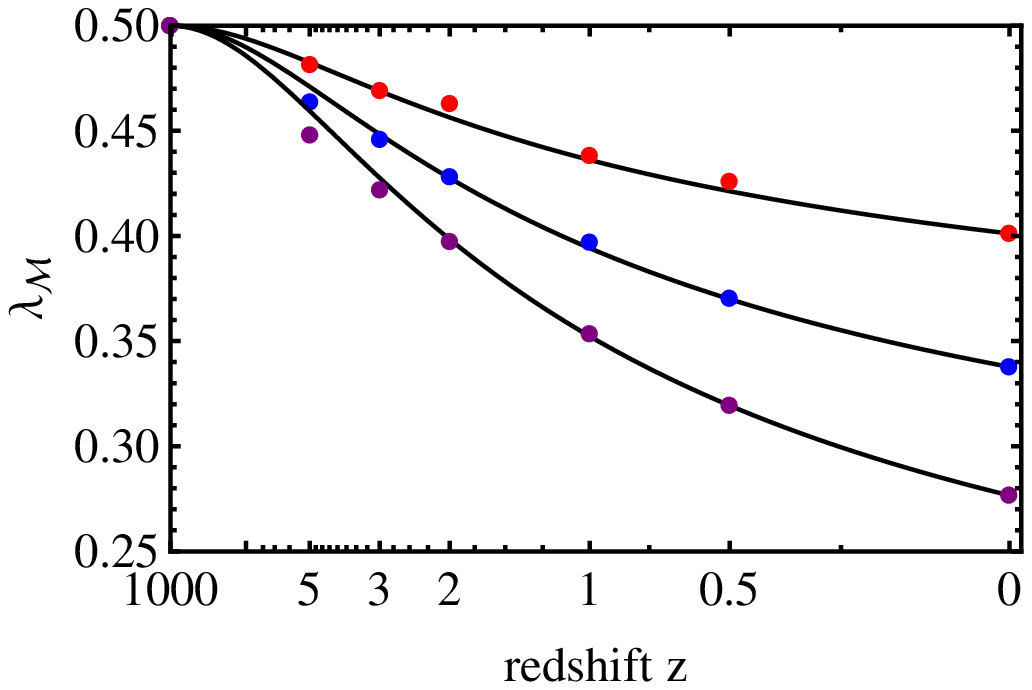}

\caption{Top left: Comparison of the evolution of the parameter $\lambda_{\CM}$
calculated from the scale factors shown in Fig.~\ref{fig:Graphen01}
(straight line), with the values derived from the $N$--body simulation
(dots). The graph shows that for the $N$--body simulation the evolution
of $\lambda_{\CM}$ is faster at the beginning and slower at the end
compared to the one for our model.\protect \\
Top right: Comparison of $\lambda_{\CM}$ as calculated from Eq. (\ref{eq:det-lambda})
for $\Omega_{m}^{\CD_{0}}\approx0.03$ (straight lines) with the data
(dots) derived from the $N$--body simulation for grid sizes of $R=5,\,10,\,15,\,20,\,30,\,50\,\rm{h^{-1}Mpc}$
(from bottom to top). The only input used to generate these curves
was the last point of the $N$--body data and the value of $\Omega_{m}^{\CD_{0}}$.\\
Bottom left: Same plot as top right but with $R=5,\,10,\,15\,\rm{h^{-1}Mpc}$ only and $\Omega_{m}^{\CD_{0}}=0.018$.\\Bottom right: Same plot as top right but with $R=20,\,30,\,50\,\rm{h^{-1}Mpc}$ only and $\Omega_{m}^{\CD_{0}}=0.035$.
\label{fig:Skalierung-von-Q}}

\end{figure*}

\subsection{Comparison with other calculations\label{sub:Comparison}}

In order to judge the quality of the model presented in the previous section,
we will now compare its predictions with other calculations in the
literature.

\subsubsection{Magnitude of the initial backreaction}

We first consider the perturbative results of \cite{brown1,brown2}. There it has been shown
that, at the recombination epoch, the magnitude of the effective energy contribution of backreaction and curvature is expected to be of the order of $2 \times 10^{-8}$ on the scale of the horizon for a Hubble rate of $h=0.7$ \cite{brown2} that has to be compared to a cosmological constant contribution in the standard model of $3 \times 10^{-9}$
(as is demanded in the strong backreaction scenario below). The corresponding term in our model (on the scale of the simulation box), $\Omega_{\CQ}^{\CD}+\Omega_{\CR}^{\CD}$ becomes, for the special parameters of the previous section (the $1/a_\CF$ scaling scenario and at an initial time where $a_{\CD}=1$), $8\times 10^{-7}$ compared to the matter density parameter. This comparison illustrates, that even if the Early Universe is in a near to homogeneous state, where one can apply perturbation theory, the Late Universe can look strongly different by the effect of the leading order perturbative $a_{\CF}^{-1}$--mode alone. When higher order --growing-- terms of their Laurent series play an important role, or when nonperturbative terms enter due to a substantial injection of backreaction at the stage of nonlinear structure formation, the situation will even become more different. As this comparison shows, the initial amplitude needed is still too large in the  $1/a_\CF$ scaling model. However, our value was measured on the scale of structure, and may still drop by going to the horizon scale, which would reduce the discrepancy.

\subsubsection{Structure formation in the concordance model}

The second comparison we want to present is the one for the evolution
of $\lambda_{\CM}$.%

In the corresponding figure, Fig.~\ref{fig:Skalierung-von-Q}, the line
shows $\lambda_{\CM}\left(z\right)$ as calculated from our model, where we have chosen to present the results in terms of $z$ to provide a more intuitive picture of the actual evolution of structure formation. The points result from our analysis of the $N$--body simulation \cite{n-body}
using a simple separation into blocks described in Appendix \ref{sub:A-simple-mesh}.
The figure shows a clear discrepancy between the Newtonian $N$--body
simulation and our averaged model. This is not surprising as we are
approximating the whole Laurent series of $\CQ_{\CF}$ and $\averageF{\CR}$
by their leading terms only. We will see in Section~\ref{sec:Power-series}
how the higher terms are connected to the shape of $\lambda_{\CM}\left(z\right)$.
Of course it may also be that a genuine relativistic simulation might
add corrections to the values of $\lambda_{\CM}$, but for the
moment we have to work with the models we have. Therefore, we will
present, in Section~\ref{sec:Modelling-structure-formation}, the characteristic features of a model that describes structure formation and accelerated expansion in the present
framework of averaged models in a consistent manner. 

\subsubsection{Structure formation in general $a_{\CF}^{-1}$--scaling models\label{sub:baryons-only}}

We finally want to report on an observation that might become of interest
in the upcoming reinterpretation of the observational data in the
backreaction context. As described in Appendix \ref{sub:A-simple-mesh},
we conducted the analysis of the $N$--body data using blocks of various
side lengths. In addition to the grid size of $5\rm{h^{-1}Mpc}$,
being at the basis of the result of the previous section, we also
used spacings of $10-50\rm{h^{-1}Mpc}$. The resulting values
for $\lambda_{\CM}\left(z\right)$ are shown as dots in the graph
on the right--hand side of Fig.~\ref{fig:Skalierung-von-Q}. As we
have seen on the left--hand side of this same figure, the $a_{\CF}^{-1}$--scaling
model using the concordance value of $\Omega_{m}^{\CD_{0}}\approx0.27$
is not able to reproduce the structure formation history as observed in the $N$--body
simulation. We may however try to derive the evolution of $\lambda_{\CM}\left(z\right)$
for different choices of $\Omega_{m}^{\CD_{0}}$. The differential
equation determining $\lambda_{\CM}\left(z\right)$ may be derived
from the definition of the latter and reads:
\begin{equation}
\frac{1}{3}\frac{a_{\CD}}{\lambda_{\CM}}\partial_{a_{\CD}}\lambda_{\CM}=\frac{H_{\CM}\left(a_{\CD},\lambda_{\CM}\right)}{H_{\CD}\left(a_{\CD},\lambda_{\CM}\right)}-1\;,
\label{eq:det-lambda}
\end{equation}
where $H_{\CD}=\lambda_{\CM}H_{\CM}+\left(1-\lambda_{\CM}\right)H_{\CE}$
and $H_{\CM}\left(a_{\CD},\lambda_{\CM}\right)$, respectively. $H_{\CE}\left(a_{\CD},\lambda_{\CM}\right)$
may be found by replacing $a_{\CM}\rightarrow a_{\CD}\lambda_{\CM}$
and $a_{\CE}\rightarrow a_{\CD}\left(1-\lambda_{\CM}\right)$ in Equation
(\ref{eq:Int-M-Ham}). Using the relations between the different
parameters discussed in Subsection~\ref{sub:Free-parameters}, we
may numerically solve (\ref{eq:det-lambda}) after having specified
the two remaining parameters $\Omega_{m}^{\CD_{0}}$ and $\lambda_{\CM_{0}}$.
The result for $\Omega_{m}^{\CD_{0}}\approx0.03$ is shown by the
lines in the graph on the right--hand side of Fig.~\ref{fig:Skalierung-von-Q}.
There we have fixed, for each curve $\lambda_{\CM}\left(z;\Omega_{m}^{\CD_{0}},\lambda_{\CM_{0}}\right)$,
$\lambda_{\CM_{0}}$ to the end value of the result from the
$N$--body simulation. It is interesting to notice, that the common value
of $\Omega_{m}^{\CD_{0}}\approx0.03$ allows us to describe the evolution
on diverse scales quite well and that the shape of the function as
defined by Equation (\ref{eq:det-lambda}) meets the actual form without
having to use the theoretically unmotivated fitting ansatz that will
be presented in Subsection~\ref{sub:Quantitative-conclusions}. The
fit may even be improved if one allows for a weak scale dependence
of $\Omega_{m}^{\CD_{0}}$, as for $\Omega_{m}^{\CD_{0}}\approx0.018$
the small scales are fitted in an exquisite manner and for $\Omega_{m}^{\CD_{0}}\approx0.035$
this is true for the large scale evolution.

If this is just a coincidence or if this points to a deeper physical interpretation of
the backreaction effect in terms of its capability to unify Dark Energy and Dark Matter in one
effective fluid cannot be finally decided from this investigation. 
It is clear that the backreaction terms imply the emergence of both dark components,
however, in an entangled way that does not allow us to assign Dark Matter to over--dense and Dark Energy to underdense regions uniquely due to the changing variance between the two types of regions. 
We may speculate that, in the course of a reinterpretation
of observations in the context of the averaged equations, the role of
Dark Matter in structure formation on large and intermediate scales might be attributed to average
curvature and backreaction components that effectively yield the same
distribution of visible matter as CDM--models (but see the constraining remarks of Sec.~\ref{subsub:dark-matter}). But it may also be that
a low value of $\Omega_{m}^{\CD_{0}}\approx0.03$ turns out to be incompatible
with other observations. These questions will be addressed in upcoming
work when testing the $a_{\CF}^{-1}$--scaling model against observations
as done for a simple $a_{\CD}^{-1}$--model in \cite{morphon:obs} (with an improved template metric).
Apart from observational issues the model has to be compared to N--body simulations with baryonic and neutrino
content only in order to quantitatively support this necessarily speculative aspect of our model. 
A naive interpretation of the low--$\Omega_{m}^{\CD_{0}}$ model also runs into the same difficulties of high initial amplitudes of perturbations as in the old
studies of baryonic universe models. Furthermore, since independent evidence for Dark Matter exists \cite{roos}, especially on small scales and unlike
the situation for Dark Energy, we cannot conclude from this work that Dark Matter may be fully identified
with backreaction effects. But, our study clearly indicates the need to exploit this aspect of the backreaction effect.

\subsection{Discussion}

It may be helpful to hold in for a moment and put into perspective what we have seen in
this section. It has been shown that in a model that possesses
the $a_{\CF}^{-1}$--limiting behavior of the backreaction components
and that starts from almost homogeneous
initial conditions as in the standard model, the simple fact that there is a lot of structure
today in terms of volume dominance of devoid regions (manifesting itself in our model through a value of $\lambda_{\CM}$ around
$0.1$) implies some $\Lambda$--like accelerated expansion of
the volume scale factor $a_{\CD}$. Even though this model is too
basic as is indicated by the discrepancy seen in Fig.~\ref{fig:Skalierung-von-Q},
it clearly illustrates the way in which one would expect backreaction
to act. (Here we do not want to over--emphasize the result that a pure baryonic matter content would even do the job.)  
To avoid the impression that we just replaced the mysterious
Dark Energy component by some other mysterious component, i.e. the backreaction
term, let us emphasize the different physical situation. In the general relativistic
framework that lies at the basis of the averaged equations
the emergence of structure is associated with a geometrical deformation of the underlying
spacetime. In the comoving time--synchronous slicing chosen to describe the
dust universe of \ref{sub:The-average-Einstein}, there is no movement
of matter particles with respect to the space. All inhomogeneities
emerge from the distortions of space itself. This results in the
emergence of intrinsic curvature of the spatial hypersurface, reflected
in the three Ricci scalar $\CR$, as well as extrinsic curvature of
this same hypersurface. This extrinsic curvature is what makes up the
backreaction as it has been shown that it may be defined instead of
kinematical quantities as in Eq. (\ref{eq:Def-Q}), in terms of invariants
of the extrinsic scalar curvature (see \cite{buchert:dust}). To gain
some intuition it may be helpful to see Equation
(\ref{eq:Raychaudhuri-Mittel}) not in an active sense stating that
there is some fundamental component that forces the scale
factor to accelerate, but rather in a passive sense that it traces
the complex evolution resulting from the full Einstein equations with
the only active component being the perfect fluid dust matter. Then
one may picture the evolution to be governed by gravitational instability
that causes inhomogeneities to grow, which manifest itself in extrinsic curvature
showing up in the backreaction term, but also -- and this is the key issue -- in intrinsic curvature of space 
in which structures emerge.  How acceleration in this context
may be understood will be discussed in Appendix \ref{sub:A-first-example}.

In this general--relativistic picture one may also see that, even without a cosmological constant, we have some kind of "Dark Energy" in our model universe
that now has a clear physical interpretation: It is the curvature energy of the spatial hypersurface, communicated by an effective potential energy in the morphon field correspondence of backreaction \cite{morphon}.
This new kind of "Dark Energy" therefore necessarily emerges when structure forms and it is this "Dark Energy" the title of this paper is referring to.

Having emphasized the role of intrinsic curvature one cannot argue that this picture should yield the same results
as the Newton--inspired picture of structure forming in space and
matter moving around. If the averaged intrinsic curvature evolves differently as compared with a
constant curvature model -- and this is a generic outcome of the fact that the kinematical backreaction term and the averaged scalar curvature are
coupled -- then curvature plays a substantial role for the structure formation history, but also 
for the interpretation of observational data. While metrical deviations from a flat space may be small, derivatives of the metric
can be substantial, as has recently
been shown by estimating its magnitude in \cite{estim}. In addition
the existence of a strong average scalar curvature term today is not excluded
by the data. A study on light propagation in
statistically homogeneous universes \cite{rasanen:light} has shown,
that the position of the first acoustic peak in the CMB spectrum is
consistent with a non--negligible amount of curvature if the expansion
history of the scale factor $a_{\CD}$ follows a $\Lambda$CDM--evolution (see also \cite{rasanen:light2}).
As this condition is fulfilled in our model, the occurrence of $\Omega_{\CR}^{\CD}\cong 1$
today is not problematic, if observational data are interpreted in the backreaction
context.

\section{Modeling structure formation and accelerated expansion: strong backreaction scenario\label{sec:Modelling-structure-formation}}

After having explored the partitioning approach with the help of a simple
$a_{\CF}^{-1}$--scaling solution in the previous section, which
provided the result that this leading perturbative mode alone is able
to account for a $\Lambda$--like accelerated expansion if we insert
what we know about the matter content and the structures of today's
Universe, it may be interesting to learn how the higher terms in the
Laurent series of $\CQ_{\CF}$ and $\averageF{\CR}$ have to contribute
to give rise to the evolution of structures as derived from the $N$--body
simulation. In other words we are searching for the nonperturbative
behavior of $\CQ_{\CF}$ and $\averageF{\CR}$ that will match the simulated
structures. One would like to get a handle on the expansion behavior
from simply tracing this evolution with the help of $\lambda_{\CM}\left(z\right)$
but as the system will still not close one has to find another condition
to come to results not being based on assumptions on the scaling of
the unknown backreaction and curvature terms themselves. Even if, in
principle, all the quantities that figure in the averaged equations
are measurable, this is very difficult in practice and we do not yet
have a precise idea how they evolve. For the backreaction and curvature terms, Ref.~\cite{buchertcarfora}
provides strategies on how they could be determined, but there are no quantitative results yet.

We will therefore only show here that the averaged equations are able
to describe accelerated expansion and structure formation in a
model using the partitioning introduced in Section~\ref{sec:Volume-partitions}.
The philosophy is that we want to show for which evolution of the regional backreaction and curvature terms the phenomenological global $\Lambda$CDM
model, that provides a good fit to a large number of data, 
may be understood to be the result of the evolution of the
physical variables extrinsic and intrinsic curvature.

\begin{figure*}
\includegraphics[width=0.47\textwidth]{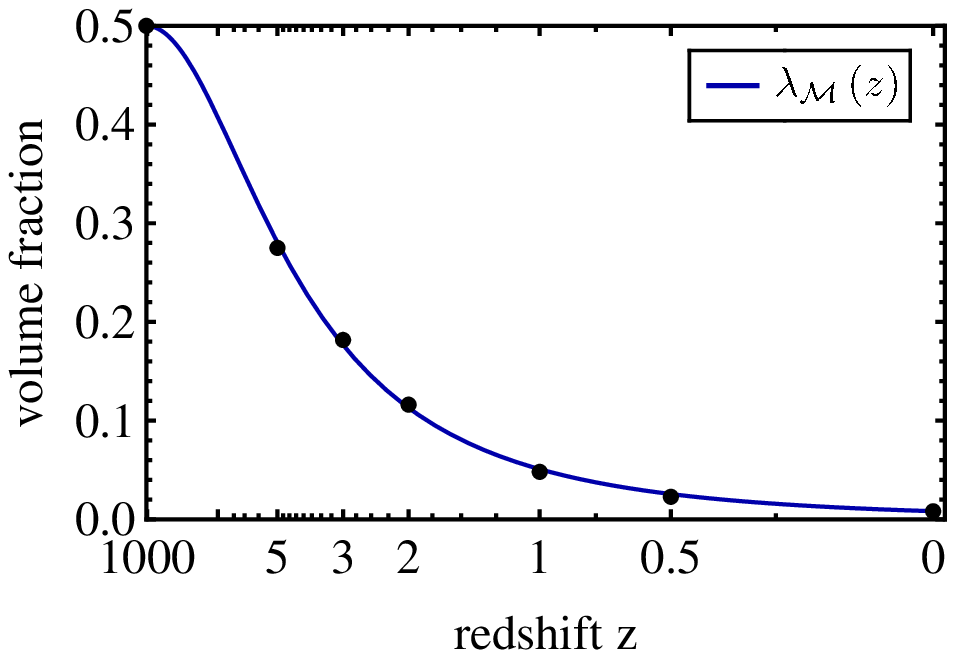}%
\hspace{0.05\textwidth}%
\includegraphics[width=0.47\textwidth]{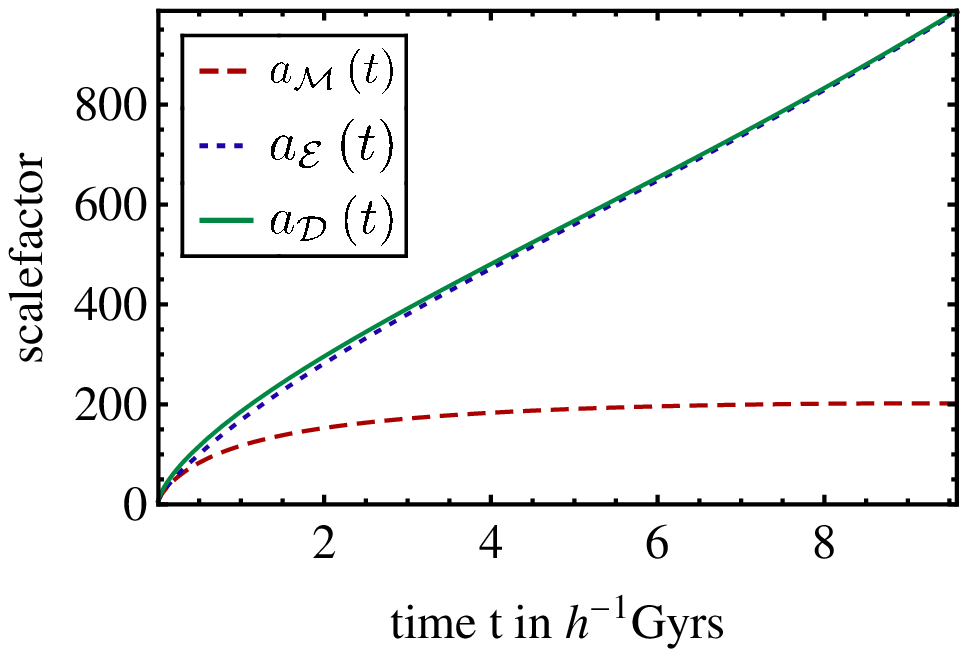}

\vspace{0.05\textwidth}%
\includegraphics[width=0.47\textwidth]{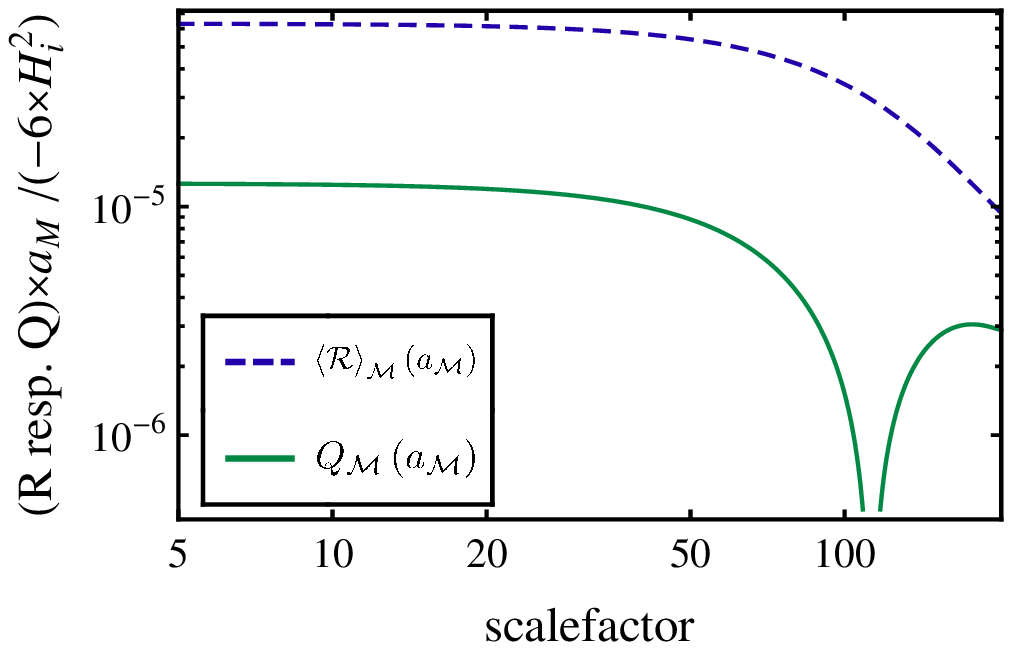}%
\hspace{0.05\textwidth}%
\includegraphics[width=0.47\textwidth]{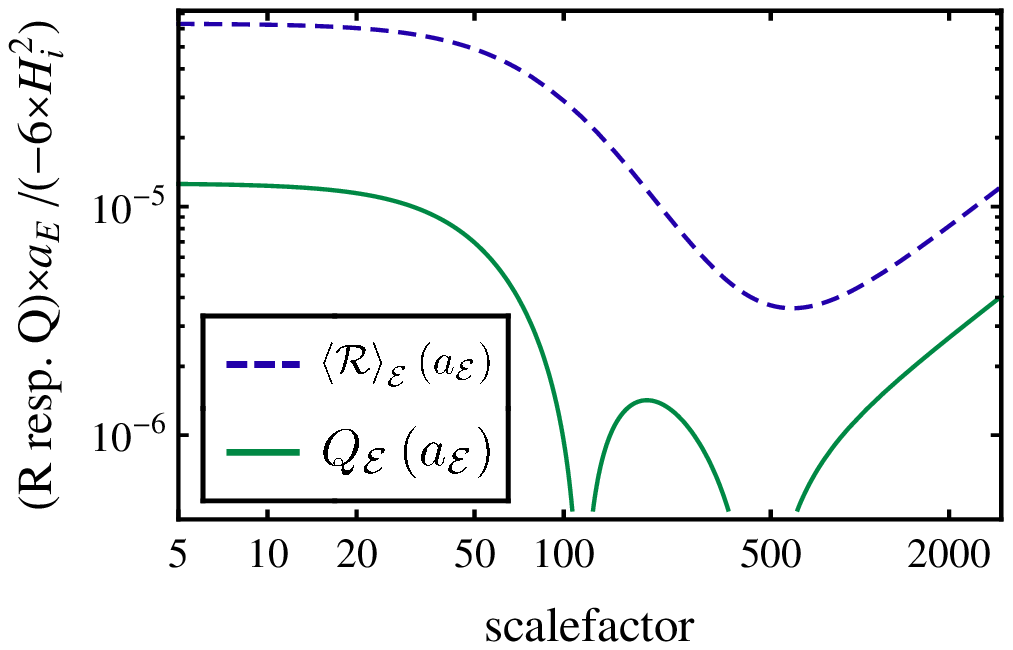}

\caption{Two partition model for a $\Lambda$CDM behavior of the $\CD$ scale
factor $a_{\CD}$ and the structure formation inferred from the $N$--body
simulation. Left on top the fit of $\lambda_{\CM}$ to the data points
of the Voronoi evaluation. To its right the dependence of the scale
factors on the foliation time. Bottom left, functional dependence
of backreaction and scalar curvature on $\CM$ on the corresponding
scale factor $a_{\CM}$. Because of the double logarithmic scaling
only the absolute value is shown. $\averageM{\CR}$ is positive, $\CQ_{\CM}$
starts negative and gets positive at around $a_{\CM}\approx110$.
Bottom right, the same plot for $\CQ_{\CE}$ and $\averageE{\CR}$ as
a function of $a_{\CE}$. $\averageE{\CR}$ is negative, $\CQ_{\CE}$
positive at the beginning and in the end, with a negative period between
$a_{\CE}\approx110$ and $a_{\CE}\approx500$. In both cases the terms
$\CQ_{\CF}$ and $\averageF{\CR}$ have been multiplied by $a_{\CF}$
to emphasize the initial $a_{\CF}^{-1}$--limit.\label{fig:FLRW-Fit-Lambda}}

\end{figure*}
\begin{figure*}
\includegraphics[width=0.47\textwidth]{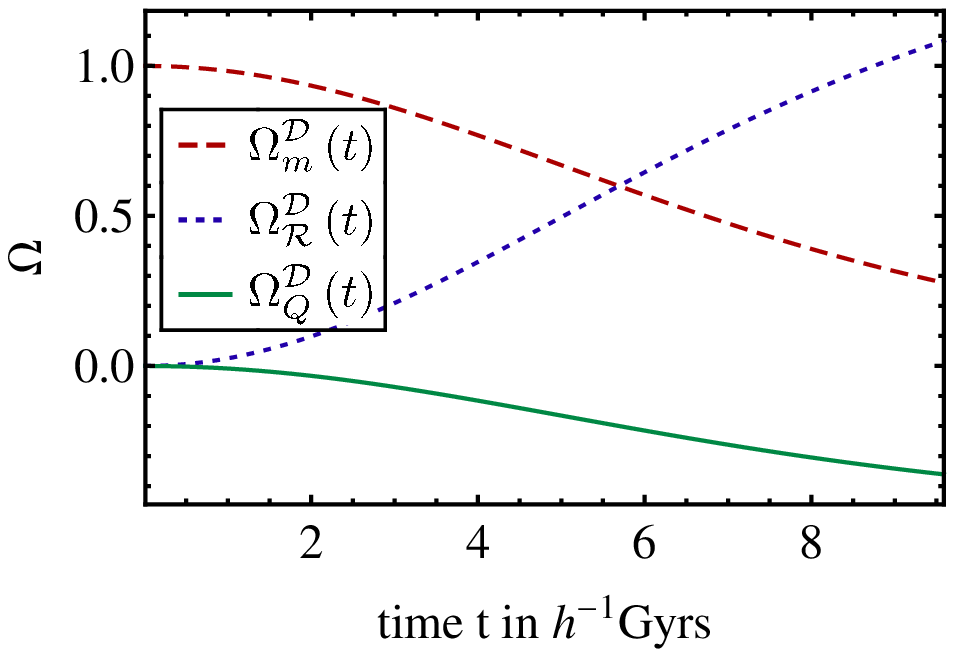}%
\hspace{0.05\textwidth}%
\includegraphics[width=0.47\textwidth]{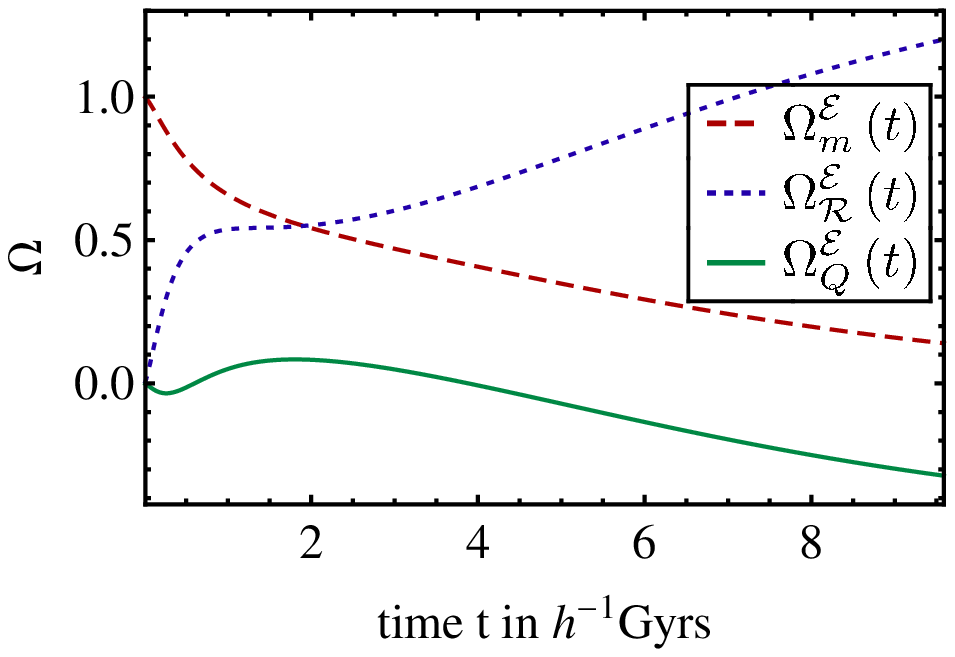}

\vspace{0.05\textwidth}%
\includegraphics[width=0.47\textwidth]{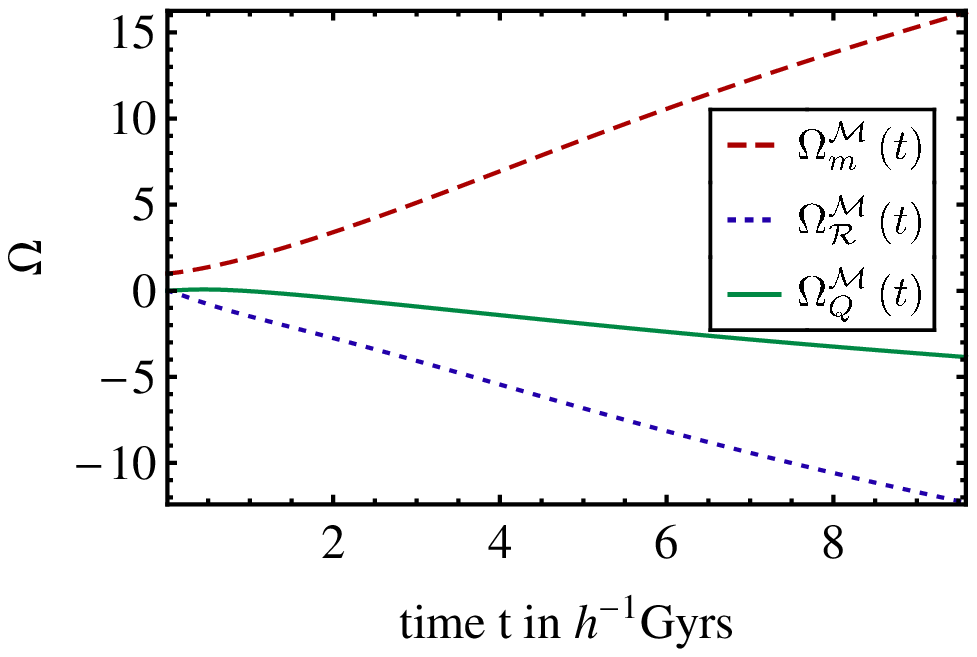}%
\hspace{0.05\textwidth}%
\includegraphics[width=0.47\textwidth]{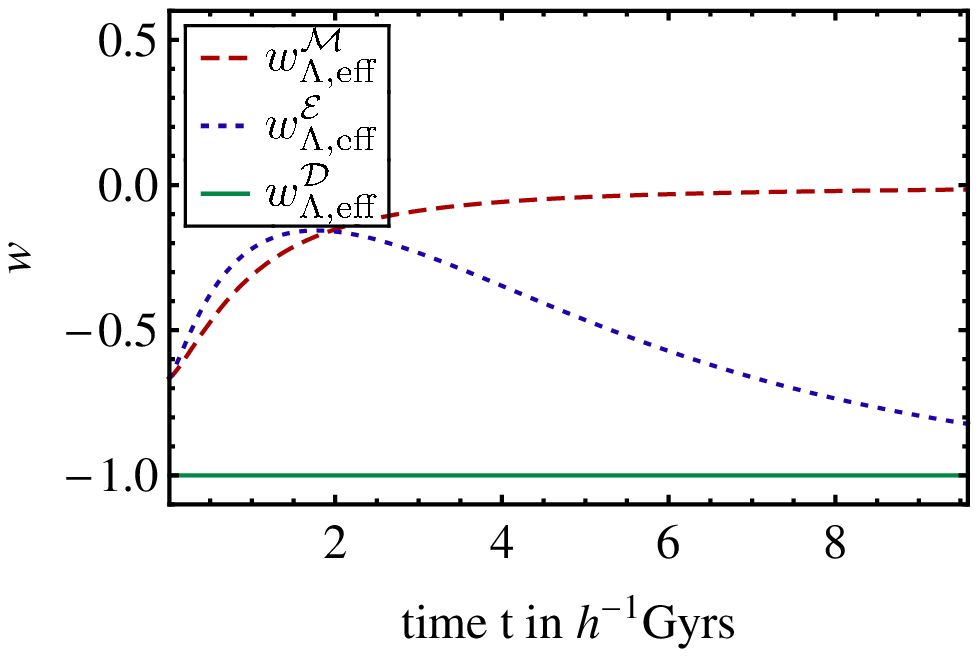}

\caption{Dimensionless parameters for the model deduced from the fit of Fig.~\ref{fig:FLRW-Fit-Lambda}. Top left the evolution on $\CD$,
on the right the one on $\CE$, left at the bottom the same
for $\CM$. Bottom right the effective equation of state of
the $X_{\mathcal{F}}$--component as defined in (\ref{eq:eq-of-state}).
The implications are discussed in Section~\ref{sub:Derived-quantities}.\label{fig:FLRW-Fit-Abg-Groessen}}

\end{figure*}

\subsection{Evolution of parameters}

In order to construct the nonperturbative model discussed above, we first
have to translate the assumptions of $\Lambda$CDM evolution and $N$--body
structure formation into quantities of our model. The $\Lambda$CDM
behavior means $\Omega_{\CQ}^{\CD}+\Omega_{\CR}^{\CD}=\Omega_{\Lambda}^{Friedmann}$,
which results in $\CQ_{\CD}\left(t\right)=\Lambda$ and $\average{\CR}=-3\Lambda$.
$\Lambda$ is determined by requiring $\Omega_{\CQ}^{\CD_{0}}+\Omega_{\CR}^{\CD_{0}}\approx0.7$.
This is possible, since the cosmological constant is a particular exact scaling solution of the averaged
equations. 
To find $\lambda_{\CM}\left(a_{\CD}\right)$ we analyzed the $N$--body
simulation \cite{n-body} using this time a separation into subvolumes
based on a Voronoi tesselation of the simulation volume. 
The resulting data
points for five different redshifts are shown in Fig.~\ref{fig:FLRW-Fit-Lambda}.
They are fitted with a functional ansatz of the form
\begin{equation}
\lambda_{\CM}\left(a_{\CD}\right):=\frac{1}{2}\frac{1}{1+\left(\frac{a_{\CD}}{\alpha_{\CM}}\right)^{2}+\left(\frac{a_{\CD}}{\beta_{\CM}}\right)^{4}}
\label{eq:Fitfunktion-Lambda}
\end{equation}
 and the best--fit parameters providing the curve of Fig.~\ref{fig:FLRW-Fit-Lambda}
were $\alpha_{\CM}\approx191$ and $\beta_{\CM}\approx419$.

\subsubsection{Results for the scaling of $\CQ_{\CF}$ and $\averageF{\CR}$}

The resulting scale factors of the domains are shown in the upper
right panel of Fig.~\ref{fig:FLRW-Fit-Lambda}. Comparing them to
the evolution of the scale factors for the $a_{\CF}^{-1}$--scaling
it becomes clear, that a faster structure formation in the $N$--body case,
which was found in Fig.~\ref{fig:Skalierung-von-Q}, requires $a_{\CM}$
to slow down much earlier. The size of the $\CM$--regions is therefore
nearly constant throughout a long period. This is again qualitatively
as expected, because the overdense regions virialize and decouple
from the overall expansion. $a_{\CE}$ evolves similar to the $a_{\CF}^{-1}$--case,
but finally has to take over the accelerated expansion and therefore
has a limiting behavior of a cosmological constant evolution. This
limiting behavior might not be realistic as it implies a constant
variance of expansion rates which is assumed to shrink as the fastest
expanding regions will dominate the volume in the late--time limit.
If there is an intermediate state providing this evolution is speculative.

The lower panels of Fig.~\ref{fig:FLRW-Fit-Lambda} show the evolution
of $\CQ_{\CM}\left(a_{\CM}\right)$ resp. $\averageM{\CR}\left(a_{\CM}\right)$
(left) and $\CQ_{\CE}\left(a_{\CE}\right)$ resp. $\averageE{\CR}\left(a_{\CE}\right)$
(right). They were multiplied by $a_{\CM}$ resp. $a_{\CE}$ to point
out the $a_{\CF}^{-1}$--behavior at the beginning of the evolution.
It is interesting that this limiting behavior, which is -- as already
mentioned -- in accord with the perturbative result of Li and Schwarz
\cite{gaugeinv,li:thesis}, arises naturally in the present setup.
The reason why this is the case will be explained in Section~\ref{sec:Power-series}.
It should be noted that at the beginning, the curvature on $\CM$
is positive, the backreaction negative. For $\CE$ it is just the
opposite. Regarding the further evolution it can be seen that the
backreaction changes sign (on $\CE$ even twice) and that $\CQ_{\CF}$
and $\averageF{\text{\ensuremath{\CR}}}$ shrink faster than $a_{\CF}^{-1}$.
The reason for this behavior can be seen in the shape of $\lambda_{\CM}$
in Fig.~\ref{fig:FLRW-Fit-Lambda}. First we need a strongly different
behavior of $\CM$-- and $\CE$--regions to assure that $\lambda_{\CM}$
shrinks rapidly. Therefore $\CQ_{\CM}$ is negative and acts like matter
to slow down the expansion of $\CM$, whereas $\CQ_{\CE}$ is positive
to lead to a faster growth. Then the growth on $\CE$ is slowed down
by a negative $\CQ_{\CE}$, whereas $\CQ_{\CM}$ becomes positive to counterbalance
the deceleration of the matter component to lead to the nearly constant
part of the evolution of the $a_{\CM}$--scale factor of Fig.~\ref{fig:FLRW-Fit-Lambda}
upper right. The linear rise for $\CE$ at the end encodes the cosmological
constant behavior, meaning that $\CQ_{\CE}\left(a_{\CE}\right)=-\frac{1}{3}\averageE{\CR}\left(a_{\CE}\right)=\rm{const.}$,
which is reflected by an increasing line in this figure as we are multiplying
it by $a_{\CE}$.

\subsubsection{Derived quantities\label{sub:Derived-quantities}}

Figure~\ref{fig:FLRW-Fit-Abg-Groessen} finally shows the functional
form of the dimensionless parameters. The upper left graph simply
shows the evolution imposed by the required cosmological constant
behavior of $\CD$ going from the matter dominated era $\Omega_{m}^{\CD_{i}}=1$
to today's value of $\Omega_{m}^{\CD_{0}}=0.27$. The rest, which is
dubbed Dark Energy in the standard concordance model is now
represented by our $X$--matter and split into backreaction and curvature.
The $\CE$--regions in the upper right graph develop rapidly {}``strong''
negative curvature, which is however not due to a rise in the $\averageE{\CR}$--term
(as may be seen from Fig.~\ref{fig:FLRW-Fit-Lambda}), but to the faster
decrease of the $\averageE{\varrho}$--contribution. The evolution
on $\CM$ shows that one has to be careful in the interpretation of
the $\Omega$--parameters. Due to the approximately constant scale
factor $a_{\CM}$, which results in a nearly constant $\averageM{\varrho}$,
the division by $H_{\CD}$ which is decreasing makes $\Omega_{m}^{\CM}$
increase rather strongly. If one interprets $3H_{\CD_{0}}^{2}/\left(8\pi G\right)$
as the critical density, this increase in $\Omega_{m}^{\CM}$ reflects
the emergence of a strong density contrast between the $\CM$--regions
and the averaged universe model.

\begin{figure*}
\includegraphics[width=0.415\textwidth]{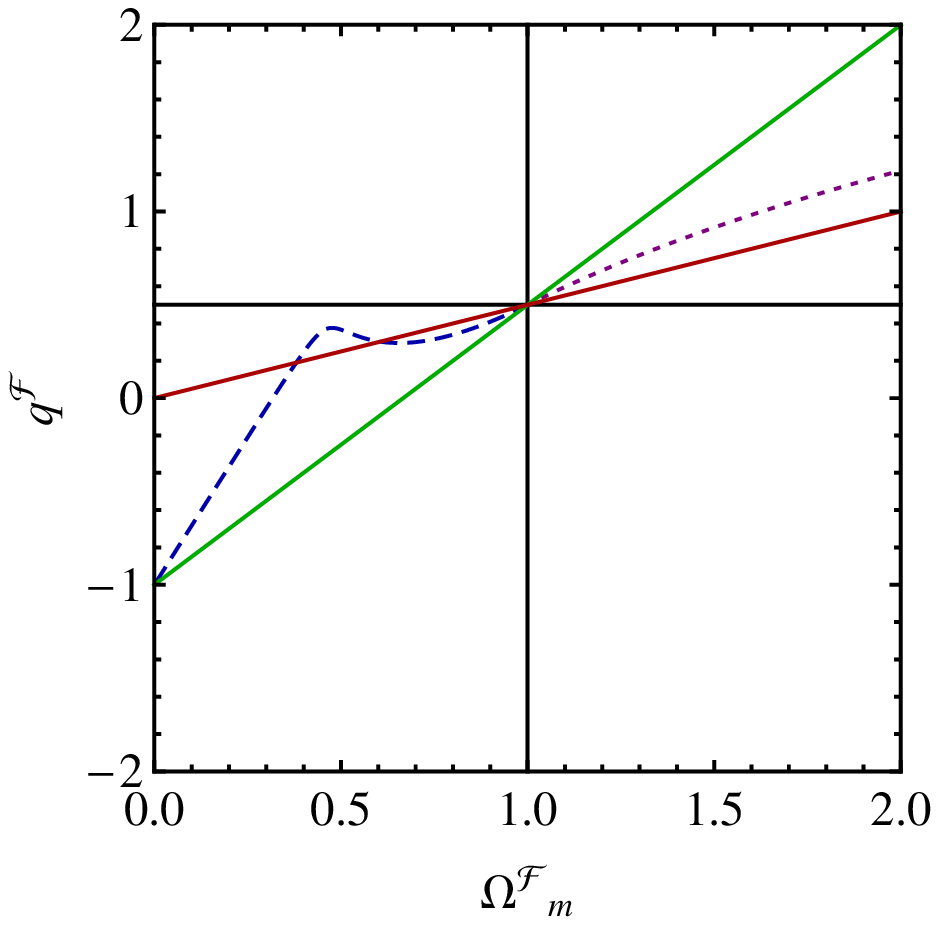}%
\hspace{0.05\textwidth}%
\includegraphics[width=0.525\textwidth]{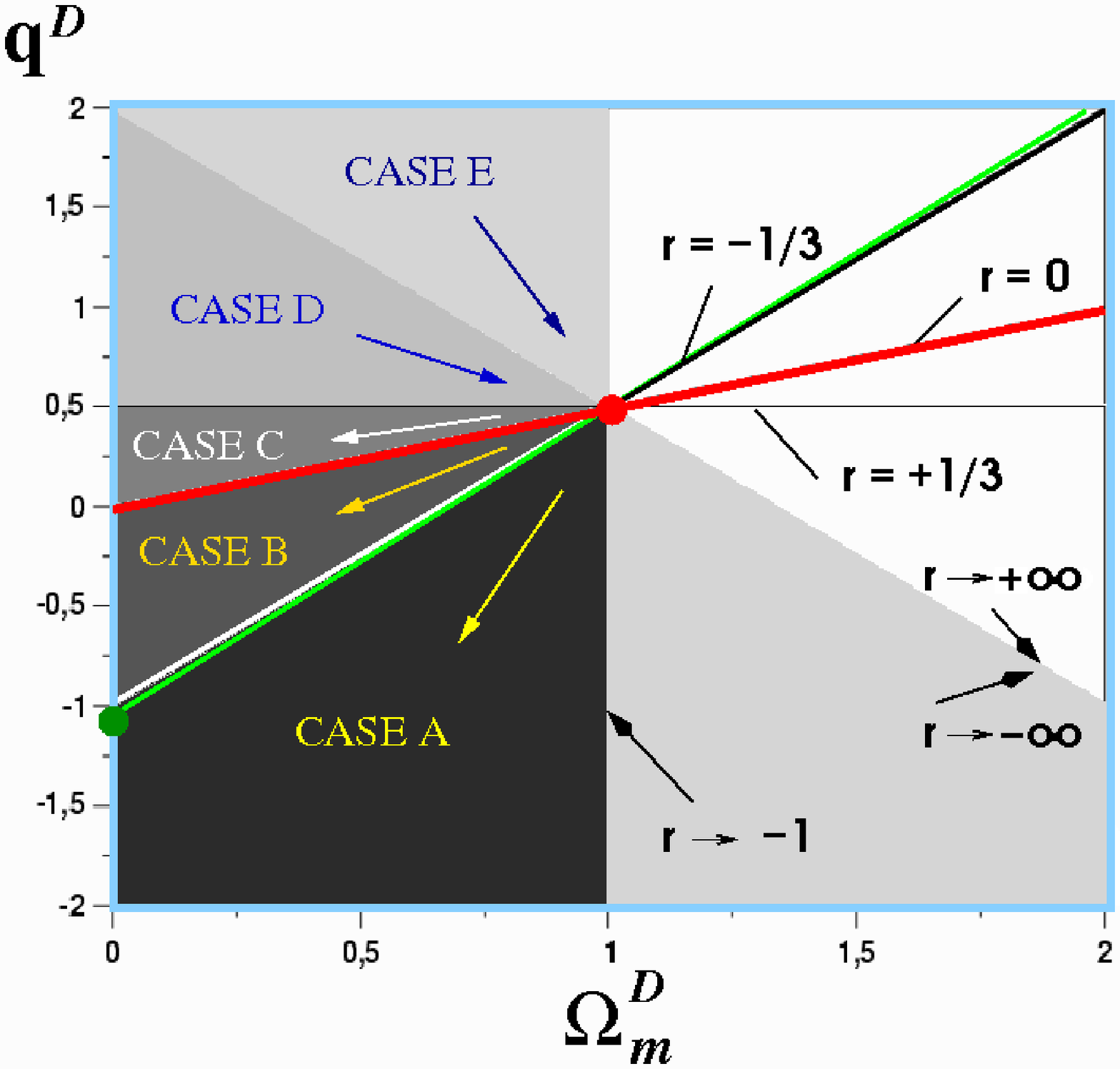}

\caption{{}``Cosmic phase space'' of the solutions of the averaged equations.
On the right  the schematic partition discussed in \cite{morphon}.
On the left the curves for $\CM$ (dotted) and $\CE$ (dashed) regions
derived from the $N$--body simulations. It should be noted that they
are shown here in the same plot only for economic reasons and actually
live in two different phase spaces.\label{fig:Skalenraum}}

\end{figure*}

Another interesting plot is shown in Fig.~\ref{fig:FLRW-Fit-Abg-Groessen}
in the lower right panel. It presents the effective equation of state
of the $X$--matter component of our model composed of $\CQ_{\CF}$ and
$\averageF{\CR}$ as $X_{\CF}:=\CQ_{\CF}+\averageF{\CR}$. $w_{\Lambda,\rm{eff}}^{\CD}=-1$
reflects the imposed cosmological constant behavior on $\CD$, whereas
the initial values on $\CE$ and $\CM$, $w_{\Lambda,\rm{eff}}^{\CM}=w_{\Lambda,\rm{eff}}^{\CE}=-2/3$,
are related to the $a_{\CF}^{-1}$--limit shown in Fig.~\ref{fig:FLRW-Fit-Lambda}.
$w_{\Lambda,\rm{eff}}^{\CE}$ approaches $-1$ and $w_{\Lambda,\rm{eff}}^{\CM}$
rapidly goes to zero. This means that it is scaling like a matter
contribution with $a_{\CM}^{-3}$ in the end, but as the sign of $\CQ_{\CM}$
is positive, it acts as {}``repulsive matter'', counterbalancing
the deceleration of the matter component to lead to a nearly constant
$a_{\CM}$. 
Again, this is expected physically. While our model does not describe the relevant small--scale
effects like velocity dispersion 
and rotation, effects that stabilize the virialized regions, the N--body simulation does describe 
these effects.  The evolution of the $\CQ_{\CM}$--term thus rather reflects the behavior of a more general backreaction term as expressed in \cite{buchert:fluid}.

\subsection{Cosmic phase space}

Another possibility to characterize the evolution of $\CQ_{\CF}\left(a_{\CF}\right)$
and $\averageF{\CR}\left(a_{\CF}\right)$ is to trace their solutions in a {}``phase
space'' introduced in \cite{morphon}. Its dimension is two, i.e. there is room for
many homogeneous, almost--isotropic effective states, while the homogeneous--isotropic
solutions just form a line in this space.
Every curve in this space
represents a solution of the averaged equations, while straight lines represent the 
class of scaling solutions. It is shown schematically
on the right--hand side of Fig.~\ref{fig:Skalenraum}. The different
sections are exhaustively discussed in \cite{morphon}. We will repeat
what is necessary for our purpose in the following. The coordinates
of this space are chosen to be the matter parameter $\Omega_{m}^{\CF}$ and the
deceleration parameter $q^{\CF}$, defined as
\begin{equation}
q^{\CF}:=-\frac{\ddot{a}_{\CF}}{a_{\CF}H_{\CF}^{2}}=\frac{H_{\CD}^{2}}{H_{\CF}^{2}}\left[\frac{1}{2}\Omega_{m}^{\CF}+2\Omega_{\CQ}^{\CF}\right]\;.
\end{equation}
This latter was chosen instead of $\Omega_{\CQ}^{\CF}$ alone in order to have
an additional intuitive meaning. In this space, every straight line
passing through the center, which represents the EdS model, corresponds
to an elementary scaling solution $a_{\CF}^{-n}$. It has been shown
in \cite{morphon} that the EdS model $\left(\Omega_{m}^{\CF},q^{\CF}\right)=\left(1,1/2\right)$
is in addition a saddle point for the dynamics of a universe model described
by the averaged equations. Even a small initial backreaction -- which
is always present due to the observed emergence of structure -- will
drive the expansion away from it and will result in accelerated expansion
if the deviation goes in the corresponding sector, which is the one
on the lower left--hand side. Besides the EdS model, further special
solutions are the line $r=1/3$ ($r$ has been introduced in Eq. (\ref{eq:Prop-Index-Skalierung})),
corresponding to models with Friedmannian kinematics but rescaled
cosmological parameters, and $r=0$ representing models without backreaction.
In this case curvature reduces to a constant curvature $a_{\CF}^{-2}$
behavior and they are therefore scale--dependent Friedmannian models.
The line $r=-1/3$ comprises models for which backreaction acts like
a scale--dependent cosmological constant. The introduction of a genuine
cosmological constant would simply shift the diagram down by its value
if it is positive.%

Figure~\ref{fig:Skalenraum} shows the form of our solutions in this
space. The dashed curve is the one for the $\CE$--regions, the dotted
one is for $\CM$. Both begin at the EdS model in the center as they
are initially matter dominated. Because of their $a_{\CF}^{-1}$--limit
they lie on the line corresponding to $r=-1/5$. Their opposite signs
that are responsible for the vanishing initial backreaction on $\CD$
make them evolve from the center in different directions. Both lines
get shallower and $\CE$ approaches the line with Friedmannian kinematics
$\left(r=1/3\right)$ until structure formation sets in and the growing
variance of expansion rates drives it to the line of a scale--dependent
cosmological constant $\left(r=-1/3\right)$. The deceleration parameter
is for $\CM$ not the best quantity as it is not guaranteed that it
does not diverge for $\dot{a}_{\CF}=0$.

\section{Power series of the backreaction terms\label{sec:Power-series}}

\subsection{Calculation}

In Section~\ref{sec:Modelling-structure-formation} we have seen
that the backreaction and curvature terms show an initial $a_{\CF}^{-1}$--behavior.
In the following we will explore where this comes from and rederive this behavior
in our present context of the partitioning into two subregions. We
first present the result in the form of a proposition:

\paragraph*{Proposition}

For every evolving spacetime that possesses a flat Einstein--de Sitter
limit at the beginning of its evolution, the backreaction and curvature
dependence on the volume scale factor $a_{\CF}$ on arbitrary subregions
$\CF$ stemming from a metric--compatible partition of compact domains in the hypersurfaces of 
constant time $\CD$ and that are evolving differently from one another,
may be expressed as a Laurent series beginning with $a_{\CF}^{-1}$
for backreaction and with $a_{\CF}^{-2}$ for curvature.

To prove this proposition we will have to find the expressions for
$\CQ_{\CF}\left(a_{\CF}\right)$ and $\averageF{\CR}\left(a_{\CF}\right)$.
They can be derived from (\ref{eq:Raychaudhuri-Mittel}) and (\ref{eq:Hamilton-Mittel})
which lead to 
\begin{eqnarray}
Q_{\CM}&=&\frac{2}{3}\frac{\lambda_{\CM_{i}}k^{3}}{a_{\CM}^{3}\left(t\right)}+3\frac{\ddot{a}_{\CM}\left(t\right)}{a_{\CM}\left(t\right)}\label{eq:Q-Stoerung}\\
\averageM{\CR}&=&2\frac{\lambda_{\CM_{i}}k^{3}}{a_{\CM}^{3}\left(t\right)}-6\frac{\dot{a}_{\CM}^{2}\left(t\right)}{a_{\CM}^{2}\left(t\right)}-3\frac{\ddot{a}_{\CM}\left(t\right)}{a_{\CM}\left(t\right)}\;,
\label{eq:R-Stoerung}
\end{eqnarray}
 where $k=\left(\frac{9}{4}a_{\CD_{0}}^{3}\Omega_{m}^{\CD_{0}}H_{\CD_{0}}^{2}\right)^{\frac{1}{3}}$.
To calculate $\CQ_{\CF}\left(a_{\CF}\right)$ and $\averageF{\CR}\left(a_{\CF}\right)$
we therefore have to derive $\dot{a}_{\CM}\left(a_{\CM}\right)$ and
$\ddot{a}_{\CM}\left(a_{\CM}\right)$. We achieve this by the
assumption of a flat matter dominated initial universe model that leads
to 
\begin{equation}
a_{\CD}\left(t\right)=kt^{\frac{2}{3}}\;;\;\dot{a}_{\CD}\left(t\right)=\frac{2}{3}\frac{k^{3/2}}{a_{\CD}^{1/2}\left(t\right)}\;;\;\ddot{a}_{\CD}\left(t\right)=-\frac{2}{9}\frac{k^{3}}{a_{\CD}^{2}\left(t\right)}\;.
\label{eq:EdS-scaling}
\end{equation}
 To connect $a_{\CM}$ to $a_{\CD}$ we use the relation $a_{\CM}=a_{\CD}\lambda_{\CM}^{\frac{1}{3}}$
and expand $\lambda_{\CM}$ with a small parameter $\alpha$ around
$\alpha=0$ which gives
\begin{eqnarray}
a_{\CM} & = & a_{\CD}\lambda_{\CM}^{\frac{1}{3}}\left(a_{\CD}\alpha\right)
\label{eq:Entwicklung-Lambda-2O}\\
 & = & \lambda_{\CM_{i}}^{\frac{1}{3}}a_{\CD}\left(1+\lambda_{\CM_{1}}a_{\CD}\alpha+\lambda_{\CM_{2}}a_{\CD}^{2}\alpha^{2}+O\left(\alpha^{3}\right)\right)\;,\nonumber 
\end{eqnarray}
 where we have defined $\lambda_{\CM_{1}}$ and $\lambda_{\CM_{2}}$
as $\lambda_{\CM_{1}}:=\lambda_{\CM}^{\prime}\left(0\right)/\left(3\lambda_{\CM_{i}}\right)$
and $\lambda_{\CM_{2}}:=\left(-2\lambda_{\CM}^{\prime2}\left(0\right)+3\lambda_{\CM}^{\prime\prime}\left(0\right)\lambda_{\CM_{i}}\right)/\left(18\lambda_{\CM_{i}}^{2}\right)$.
A prime stands for a derivative with respect to the argument and $\lambda_{\CM}\left(0\right)$
is denoted as $\lambda_{\CM_{i}}:=\lambda_{\CM}\left(0\right)$. Differentiating
(\ref{eq:Entwicklung-Lambda-2O}) with respect to time gives $\dot{a}_{\CM}\left(\dot{a}_{\CD}\right)$
and $\ddot{a}_{\CM}\left(\dot{a}_{\CD},\ddot{a}_{\CD}\right)$. Using
(\ref{eq:EdS-scaling}) we find $\dot{a}_{\CM}\left(a_{\CD}\right)$
and $\ddot{a}_{\CM}\left(a_{\CD}\right)$, which leads to the finally
necessary $\dot{a}_{\CM}\left(a_{\CM}\right)$ and $\ddot{a}_{\CM}\left(a_{\CM}\right)$
by an inversion of (\ref{eq:Entwicklung-Lambda-2O}). This provides
\begin{eqnarray}
\CQ_{\CM} & = & \frac{2}{3}\lambda_{\CM_{i}}^{\frac{1}{3}}k^{3}\left(3\lambda_{\CM_{1}}^{2}+7\lambda_{\CM_{2}}\right)\frac{\alpha^{2}}{a_{\CM}}
\label{eq:Q-Loesung-Stoerung}\\
\averageM{\CR} & = & -\frac{40}{3}\lambda_{\CM_{i}}^{\frac{2}{3}}k^{3}\lambda_{\CM_{1}}\frac{\alpha}{a_{\CM}^{2}}
\label{eq:R-Loesung-Stoerung}\\
 &  & -\frac{10}{3}\lambda_{\CM_{i}}^{\frac{1}{3}}k^{3}\left(3\lambda_{\CM_{1}}^{2}+7\lambda_{\CM_{2}}\right)\frac{\alpha^{2}}{a_{\CM}}\;,\nonumber 
\end{eqnarray}
 and therefore proves the proposition if we take into account that
the expansion (\ref{eq:Entwicklung-Lambda-2O}) may be extended to
arbitrary order. The third-- and fourth--order terms of $\CQ_{\CM}$ are
\begin{eqnarray}
 &  & -\frac{4}{3}k^{3}\left(2\lambda_{\CM_{1}}^{3}-3\lambda_{\CM_{2}}\lambda_{\CM_{1}}-9\lambda_{\CM_{3}}\right)\alpha^{3}
\label{eq:test-1}\\
 &  & \nonumber\\
 &  & 2k^{3}\left(3\lambda_{\CM_{1}}^{4}-6\lambda_{\CM_{2}}\lambda_{\CM_{1}}^{2}-4\lambda_{\CM_{3}}\lambda_{\CM_{1}}\right.\nonumber \\
 &  & \left.+\lambda_{\CM_{2}}^{2}+11\lambda_{\CM_{4}}\right)\lambda_{\CM_{i}}^{-\frac{1}{3}}\alpha^{4}a_{\CM}\;,
\label{eq:test-2}
\end{eqnarray}
 and all higher--order terms may be calculated in a straightforward
way, but have increasingly complicated coefficients. By Equation~(\ref{eq:Integrability-curv}),
$\CQ_{\CF}\left(a_{\CF}\right)$ and $\averageF{\CR}\left(a_{\CF}\right)$
are linked and the terms $a_{\CF}^{n}$ of $\averageF{\CR}\left(a_{\CF}\right)$
are the same as those for $\CQ_{\CF}\left(a_{\CF}\right)$ simply multiplied
by $r^{-1}=-\left(n+6\right)\left(n+2\right)^{-1}$. For the second--order
term this may be seen in (\ref{eq:Q-Loesung-Stoerung}) and
(\ref{eq:R-Loesung-Stoerung}) where $r^{-1}=-5$. For the third-- and
fourth--order terms above one arrives at $\averageF{\CR}\left(a_{\CF}\right)$
when multiplying the third--order term with $r^{-1}=-3$ and the fourth--order
one with $r^{-1}=-7/3$.

These results show why we had to include the condition of a different
evolution of the subregions. If they expand in the same manner, the
ratio $\lambda_{\CM}$ stays constant which means that $\lambda_{\CM_{x}}=0$
and therefore $\CQ_{\CF}=\averageF{\CR}=0$. Another way to phrase it
is, that there is no backreaction in a completely homogeneous universe.
But if there are structures developing at any scale, we will have
a nonvanishing backreaction term and it will go as $a_{\CF}^{-1}$
in a matter dominated era.

\begin{figure*}
\includegraphics[width=0.47\textwidth]{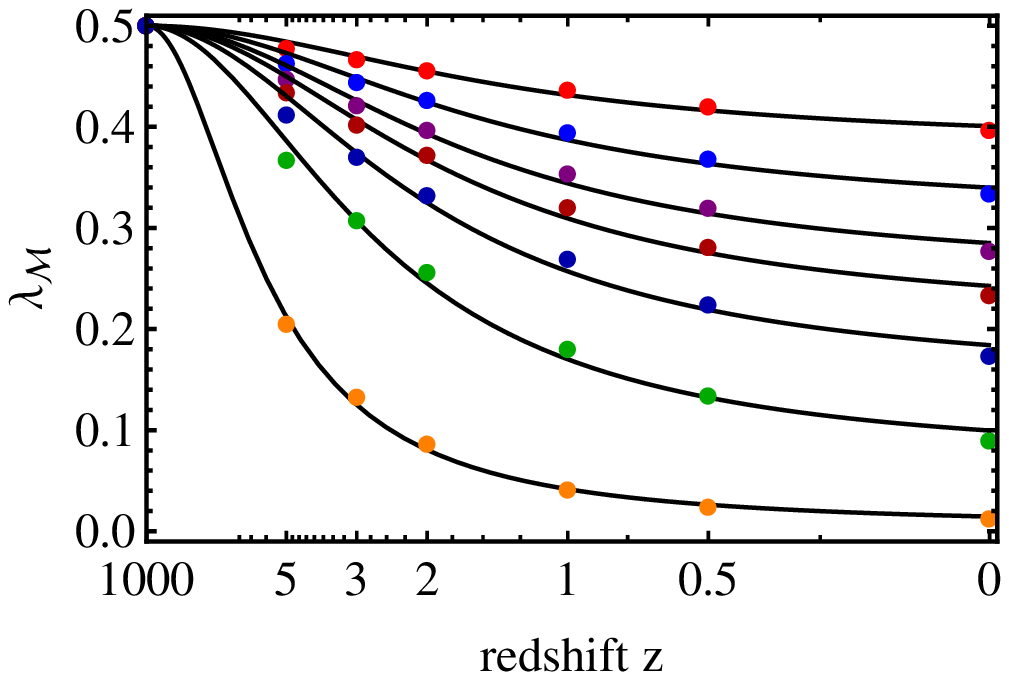}%
\hspace{0.05\textwidth}%
\includegraphics[width=0.47\textwidth]{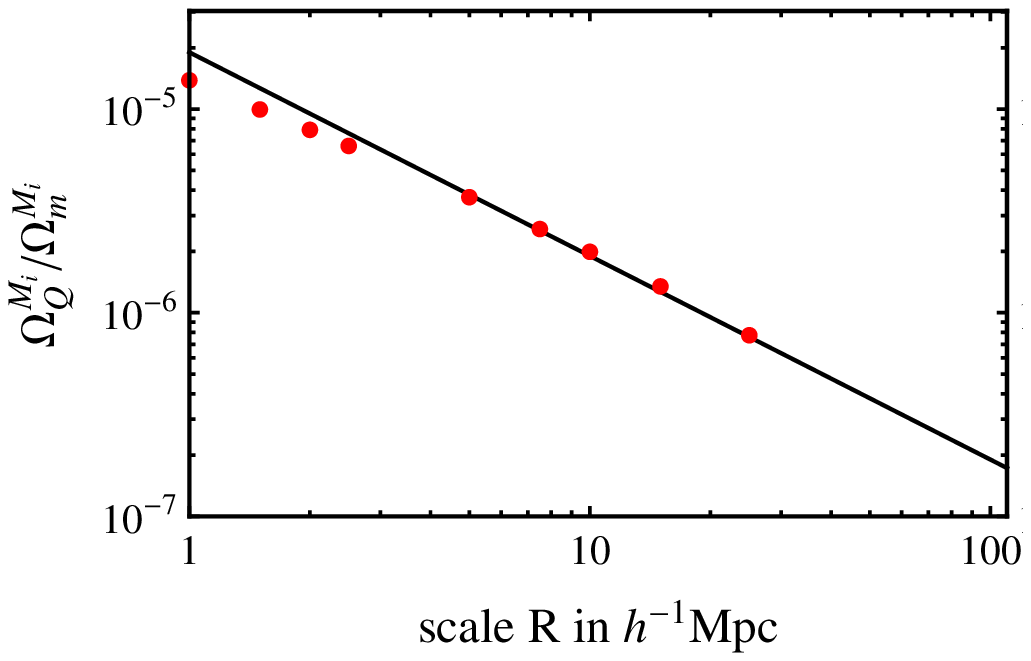}

\caption{Left: Plot of the evolution of the volume percentage of the overdense
regions $\lambda_{\CM}\left(z\right)$ for scales of $R=1,\,5,\,10,\,15,\,20,\,30,\,50\,\rm{h^{-1}Mpc}$
(from bottom to top). The data points derived by the mesh method of
Appendix \ref{sub:A-simple-mesh} from \cite{n-body} are fitted with
the functional ansatz of (\ref{eq:Fit-komplex}). Right: double--logarithmic
plot of the scale--dependence of the magnitude of the $a_{\CM}^{-1}$--term
in the Laurent series of $\CQ_{\CM}$ (\ref{eq:Q-Loesung-Stoerung})
at the initial time $t_{i}$ (the scale $R$ has been divided by two to match the widely used top--hat scales). To illustrate the trend of the data points we added a line with $R^{-1}$--scaling. The $y$--axis shows the
quantity $\Omega_{\CQ}^{\CM_{i}}/\Omega_{m}^{\CM_{\text{i}}}\approx \Omega_{\CQ}^{\CM_{i}}$, while the $x$--axis shows the top--hat grid scale in $\rm{h^{-1}Mpc}$.\label{fig:Scaling}}

\end{figure*}

\subsection{Quantitative conclusions \label{sub:Quantitative-conclusions}}

Two remarks are in order here. First we want to emphasize that the
result of Equations~(\ref{eq:Q-Loesung-Stoerung}) and (\ref{eq:R-Loesung-Stoerung})
coincides with what has been found by Li and Schwarz \cite{gaugeinv,li:thesis}
in second--order cosmic perturbation theory. They also found that on
a region $\CD$ on a flat matter dominated background the backreaction
term is of second order and scales as $a_{\CD}^{-1}$. The third--order
term is a (cosmological) constant.

Second, we want to explore what the result means quantitatively for
the initial backreaction. For simplicity let us first take the fit
to the $N$--body simulation data of Section~\ref{sec:Modelling-structure-formation}.
For the functional form (\ref{eq:Fitfunktion-Lambda}) we obtain 
\begin{equation}
a_{\CM}=\lambda_{\CM_{i}}^{\frac{1}{3}}a_{\CD}\left(1-\frac{1}{3}a_{\CD}^{2}\alpha^{2}\right)+O\left(\alpha^{4}\right)\;,
\label{eq:Entwicklung-Fitfunktion}
\end{equation}
and we may derive the expansion coefficients to be $\lambda_{\CM_{1}}=0$
and $\lambda_{\CM_{2}}=-1/3$. The first observation is that $\lambda_{\CM_{1}}=0$
causes the $a_{\CF}^{-2}$ term of the averaged scalar curvature $\averageF{\CR}$
to be zero. In view of the definition of $\lambda_{\CM_{1}}$ below
Eq. (\ref{eq:Entwicklung-Lambda-2O}) it becomes clear that this is the
case whenever $\lambda_{\CM}^{\prime}\left(0\right)=0$. This means
that, if the difference between the subregions at the beginning is
not very important so that they evolve initially in a similar way,
the term that evolves as a constant curvature term in a FLRW model
with $a_{\CF}^{-2}$ will be vanishing, provided the background does not
have a constant curvature term by itself.

If we demand that the initial matter densities are similar on all
the subregions $\Omega_{m}^{\CM_{i}}=\Omega_{m}^{\CD_{i}}$, we find
to leading order 
\begin{equation}
\frac{\Omega_{\CQ}^{\CM_{i}}}{\Omega_{m}^{\CM_{i}}}=\frac{7}{12}\alpha^{2}\;,
\label{eq:Qi-simple}
\end{equation}
which gives the initial value of the backreaction compared to the
initial matter density. This relation is interesting since it encodes
the relation between the evolution of structure in the Universe and
the initial value of the backreaction term. In (\ref{eq:Fitfunktion-Lambda})
$\alpha:=\alpha_{\CM}^{-1}$ characterizes the rate at which the
$\CM$--regions decouple from the common evolution of the subregions.
For a larger $\alpha$, $\lambda_{\CM}$ decreases more rapidly. This
faster evolution requires a bigger initial amount of backreaction
as it encodes the inhomogeneities that force the system to deviate
from the initial near to homogeneous state. The larger the inhomogeneities
are, the faster this deviation takes place.

To give a quantitative estimate we will however use a slightly different
fitting function than (\ref{eq:Fitfunktion-Lambda}) that is adopted
to the shape of $\lambda_{\CM}\left(a_{\CD}\right)$ on a larger range
of scales. We analyze to this end the $N$--body data provided by \cite{n-body}
with a mesh of varying grid size. The result are the data points of
the left graph of Fig.~\ref{fig:Scaling}. %

They are shown together with a fit using the two parameter model
\begin{equation}
\lambda_{\CM}\left(a_{\CD}\right)=\left(\beta+\frac{1/2-\beta}{1+\left(\alpha a_{\CD}\right)^{2}}\right)e^{-\frac{\alpha a_{\CD}}{\gamma\left(\alpha,\beta\right)}}\;,
\label{eq:Fit-komplex}
\end{equation}
 where $\gamma\left(\alpha,\beta\right)$ is determined by the requirement
that $\lambda_{\CM}\left(a_{\CD}\right)$ is flat for $a_{\CD}=1$.
The form of this function is motivated in Appendix \ref{sub:A-simple-mesh}.

An interesting outcome from this figure is that it confirms the idea
that $\lambda_{\CM}$ may be a good parameter characterizing the formation
of structure. This is because the different scales evolve differently
as expected from the hierarchical formation of structure. On small
scales, e.g. for $R=1\rm{h^{-1}Mpc}$, $\lambda_{\CM}$ has already
dropped to half of its initial value at a redshift of approximately
$z=6$. For scales of $R=10\rm{h^{-1}Mpc}$ this happens only
at $z\approx0.7$. Therefore, we recover at least qualitatively the
fact that structures start forming at small scales and only then assemble
to bigger ones.

Using an expansion of the form (\ref{eq:Entwicklung-Fitfunktion})
and the general form of (\ref{eq:Qi-simple}) 
\begin{equation}
\frac{\Omega_{\CQ}^{\CM_{i}}}{\Omega_{m}^{\CM_{\text{i}}}}=-\frac{1}{4}\left(3\lambda_{\CM_{1}}^{2}+7\lambda_{\CM_{2}}\right)\alpha^{2}\;,
\end{equation}
 we finally derive the initial abundances of the $a_{\CM}^{-1}$ backreaction
term that are shown on the right--hand side of Fig.~\ref{fig:Scaling}.
We have evaluated grid lengths from $1-50\rm{h^{-1}Mpc}$. The
line of Fig.~\ref{fig:Scaling} indicating an $R^{-1}$--behavior
of the scale dependence shows that in our evaluation of the $N$--body
structure formation there seems to be a discrepancy with results from
\cite{li:scale}, where the authors found that the values for
the backreaction term as derived from the power spectrum should scale
as $R^{-4}$ with the length scale $R$. In their case, however, they used a pure Harrison--Zel'dovich spectrum scaling as $k^{1}$ to derive this result. For a more realistic CDM power spectrum, the small--scale behavior is rather $k^{-2}$ which in turn means that the spatial dependence changes from $R^{-4}$ to $R^{-1}$. The region that we could evaluate with scales up to $25\rm{h^{-1}Mpc}$ (top hat), is in this small--scale regime. It would be interesting to evaluate a bigger simulation to go beyond this scale and to see the change in the scaling behavior of the backreaction term.

Quantitatively, the percentage of initial backreaction shown in Fig.~\ref{fig:Scaling} ranges from $10^{-5}-10^{-7}$ depending on the
scale (see Fig.~\ref{fig:Scaling}). These specific values show that backreaction is indeed perturbatively small in the quasihomogeneous 
epoch of the Universe. On large scales it seems insufficient to change the overall behavior as indicated by the value of $2\times10^{-8}$ on the scale of the horizon, found for example in the perturbative calculation of \cite{brown2}.
On the scale of structures, however, this perturbative contribution grows with respect to the matter content due to
its $a_{\CF}^{-1}$--scaling and is sufficient to lead to the diverse
structures we see in today's Universe.

\section{Concluding remarks and outlook\label{sec:Discussion}}

Let us discuss what we have seen in this paper. Section~\ref{sec:Volume-partitions}
showed that a consistent split of the dynamical equations governing
the averaged universe model is possible. This split also allows, in Appendix
\ref{sub:A-first-example}, to shed light on the property of
the volume scale factor to show accelerated expansion, even in a case
where there was no acceleration in the evolution of its components.
The split of the equations enabled us to construct an
averaged universe model without having to know the initial magnitude of backreaction.
Instead this latter follows from the strength of structure formation that we
put in in the form of today's volume fraction $\lambda_{\CM}$ of the
initially overdense regions $\CM$. This model implies that 
the volume scale factor $a_{\CD}$ evolves quite similarly to the
scale factor in a flat Friedmann model with about 70\% cosmological
constant. An obvious shortcoming of this model is a mismatch with
the actual time--evolution of the best--fit volume fraction with that predicted
by the $N$--body data. Surprisingly, this mismatch disappears neatly, if we reinterpret
the fundamental Dark Matter fraction in the matter parameter by 
the backreaction effect, leaving only the baryon content as fundamental. 
We do not want, however, to emphasize this result, since the Dark Matter issue
has to be investigated much beyond our simple model.

Having found that there may be more to the evolution of $\CQ_{\CF}$
than just extrapolating the perturbative $a_{\CF}^{-1}$--behavior, we investigated in Sec.~\ref{sec:Modelling-structure-formation} how to connect structure
formation and accelerated expansion within a strong backreaction scenario.
By globally imposing the particular scaling solution that mimics a cosmological constant and by assuming the structure formation history
of a Newtonian $N$--body simulation, the initial $a_{\CF}^{-1}$--limit on the subdomains is obtained, but this also provided
a nonperturbative extension of the $\CQ_{\CF}$--scaling to later
times. The emergence of the $a_{\CF}^{-1}$--scaling was then closer
investigated in Sec.~\ref{sec:Power-series} where it became clear that it
is generic for any matter dominated universe model that starts out close to
homogeneity. This also confirmed that our choice to invoke the $a_{\CF}^{-1}$--behavior
on the evolution on $\CM$ and $\CE$ instead of $\CD$ was the right
one. We could instead have identified our $\CD$--region with the one
that Li and Schwarz looked at, but Section~\ref{sec:Power-series} showed
that our $\CD$--region corresponds rather to their background. Finally,
we showed that if the Universe may be described by the average equations
and we require it to have the structure we see today, this means that the backreaction
component has to be of the order of $10^{-5}-10^{-7}$ (as compared
to the matter density and depending on the scale one is considering)
in the initial near to homogeneous phase that may be treated by perturbation
theory. In the later nonperturbative stages of the evolution of the
Universe, this tiny initial fraction is sufficient to give rise to
structures and inhomogeneities that we see in the recent epoch.

A striking result of this work is the fact that both complementary scenarios lead to 
qualitatively similar evolution laws for the backreaction terms on the largest scales, while 
the models only differ in the details, e.g. in the concrete form of the structure formation histories
and in the behavior of the variables on $\CM$--regions. Since the assumptions underlying these
two scenarios are quite orthogonal, we are confident that the $a_{\CF}^{-1}$--scaling behavior at early stages
of backreaction evolution on the subdomains (imposed in the first scenario and derived from the latter) 
should be close to the actual physical evolution that has to be confirmed by future investigations of
perturbation theory on an evolving background and by nonperturbative models that contain exact solutions
as limiting cases.

One out of a number of problems that remain is to give a clearcut interpretation of the
quantities calculated in terms of observables. As we argued it is
difficult to link the initially underdense $\CE$--regions directly
to today's voids as one might wish to. To bypass this ambiguity one
could imagine to extend the analysis to three regions $\CM$, $\CA$,
and $\CE$ where one could choose which overdensity they possess.
It would be natural to consider as $\CM$--regions the ones with overdensities
of typical clusters, the $\CE$--regions with those of voids and put
all the astrophysically unspectacular rest into the $\CA$--regions.
The advantage of probably being able to give a more refined meaning to for example
the $\Omega$--parameters is, however, counterbalanced by the inconvenience
to have to introduce another parameter in the scaling model or even
a free function $\CQ_{\CA}\left(a_{\CA}\right)$ in
the general case. First results along the lines of the scaling of Section~\ref{sec:A-simple-scaling} and the fit of Section~\ref{sec:Modelling-structure-formation}
have been obtained in \cite{wiegand}, but further analysis would
be needed to find out whether the identification is possible and if
it is therefore worthwhile to go this way.

A further refinement of the presented model would be to give up the preservation of the identity of the initially characterized subregions.
Whether this is necessary could be checked through a detailed analysis of N--body simulations. As our model is effective one could think of
implementing reaction rates between over-- and underdense regions in the spirit of nonequilibrium transitions in chemical reactions. 
A mass--action law could be devised that governs an equilibrium state between elementary subregions in which the reaction rates are equal but nonvanishing, and the transition
laws could be determined as done in nonequilibrium chemical reactions (details of such an approach in the cosmological context may be found in 
\cite{buchert:diploma}). 

We shall, however, first investigate other models that are based on relativistic Lagrangian perturbations, generalizing the Newtonian nonperturbative model investigated in \cite{bks,abundance}. This systematic attempt --
that is currently in preparation -- will provide the first generic relativistic evolution model for structure formation, including the spacetime metric for implementing observables as measured along the light cone. 

\vspace{5pt}

\begin{acknowledgments}{\small
We wish to thank Frans R. Klinkhamer for his constant interest, for encouraging us to fix the model by exploiting $N$--body data, and for critical remarks on the general setup. For valuable discussions and remarks on the manuscript we also thank Valerio Marra, Syksy R\"as\"anen, Dominik Schwarz, Henk van Elst and David Wiltshire. Finally, we would like to thank the anonymous referee for his encouragement to actually calculate the luminosity distances of our models. The work of A.W. was partially supported by the DFG under Grant No. GRK 881.}
\end{acknowledgments}

\appendix

\section{Accelerated expansion in averaged models\label{sub:A-first-example}}

We want to use this appendix to elaborate on the possibility of accelerated
expansion with local deceleration. This property of averaged models
often leads to confusion so it may be worthwhile to give
a simple example, in which one can understand intuitively its origin,
and then recap some arguments of the literature for the general case.
One popular paper that gave rise to the confusion is, for example, \cite{ishibashi}. Apart from the acceleration issue discussed in this section, it also postulates the 
negligibility of backreaction effects due to the applicability of the post--Newtonian metric. The reasons of why this argument is false, are by now given in various papers in the literature, e.g. \cite{kolb:backgrounds,estim,rasanen:perturbation}.

\subsection{A toy model}

\begin{figure}
\includegraphics[width=0.22\textwidth]{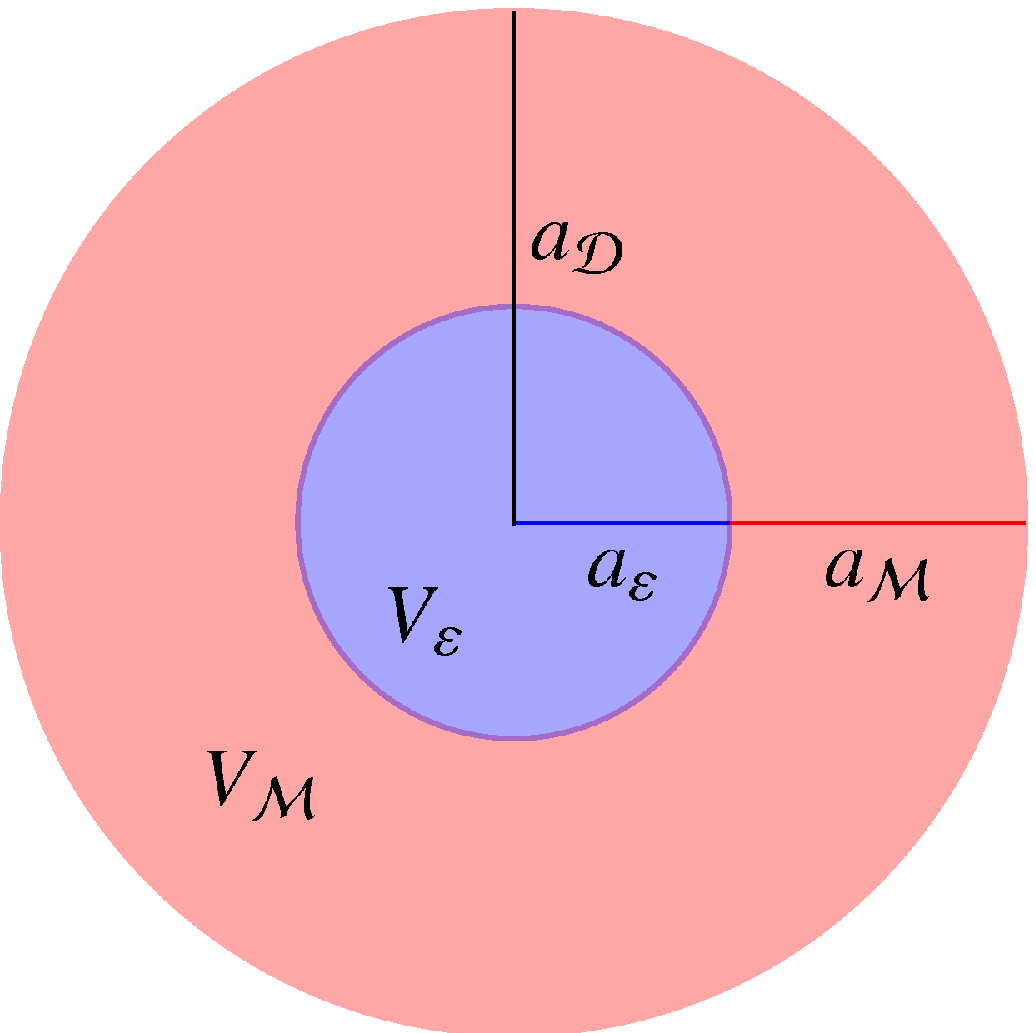}%
\hspace{0.01\textwidth}%
\includegraphics[width=0.24\textwidth]{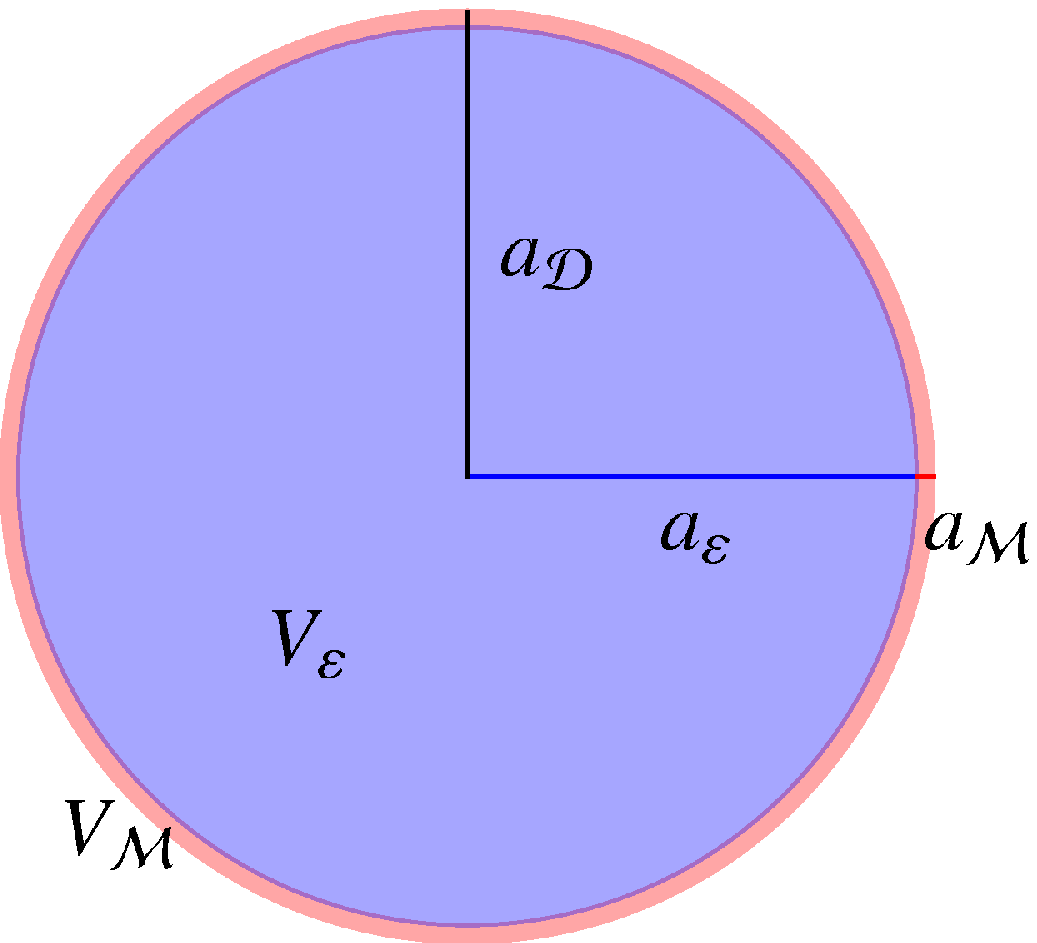}

\caption{Toy model of a two component model to illustrate accelerated
volume expansion. On the left the initial configuration
with a matterless void $\CE$ and a massive shell $\CM$. The void
in the middle expands in the sequel due to its negative curvature.
The shell does not expand any more as it contains only a virialized
matter configuration. This leads to the final state depicted on the
right where the void volume is dominating and the matter regions have
been stretched and decreased in thickness. In between, the expansion
of the total radius $a_{\CD}$ experiences an acceleration as is argued
in the text.\label{fig:Einfaches-Modell-Besch}}

\end{figure}

Let us consider the setup of Fig.~\ref{fig:Einfaches-Modell-Besch},
where we have identified two distinct regions $\CM$ and $\CE$.%

The spherical $\CE$--region in the middle is assumed to be devoid
of matter and to expand only due to its negative curvature. The shell
however is assumed to be matter dominated and virialized such that
the net expansion rate $H_{\CM}$ is zero. This means that the volume
of the $\CM$ region is constant, i.e. $\dot{a}_{\CM}=0$, and that
$\ddot{a}_{\CM}=0$. Let us consider the case when $\ddot{a}_{\CE}$
is also vanishing which leads with (\ref{eq:acc-split}) to
\begin{equation}
\frac{\ddot{a}_{\CD}}{a_{\CD}}=2\lambda_{\CM}\left(1-\lambda_{\CM}\right)H_{\CE}^{2}\;.
\label{eq:acc-without-acc}
\end{equation}
 As long as the inner region keeps expanding we will also have acceleration
of the global expansion, even if none of the two regions themselves
accelerate. In our case this can be understood as a volume effect.
As $\ddot{a}_{\CE}=0$, $\dot{a}_{\CE}$ is constant and therefore
the growth in radius within a time interval $\Delta t$ is $\Delta a_{\CE}$
and also constant. That this does not result in a constant expansion
rate for the total radius $a_{\CD}$ is due to the condition that
the volume of $\CM$ is constant. With an increasing inner radius,
the shell must become thinner to satisfy this condition. In a more
intuitive way one may think of the growth of the inner radius as if
it grew into the shell so that we had to subtract a shell with thickness
$\Delta a_{\CE}$ from the inner border of the $\CM$--shell and to
add the missing volume at the outer border of the whole configuration.
As the radius of the shell to add there is bigger than at the inner
border, we need not to add $\Delta a_{\CE}$ but a shell of a smaller
thickness to keep $\left|\CM\right|$ constant. This means that the
radius $a_{\CD}$ of the total configuration does not grow by $\Delta a_{\CE}$
but by $\Delta a_{\CD}<\Delta a_{\CE}$. With increasing size of the
expanding inner sphere and the corresponding reduction of the thickness
of the $\CM$--shell, as shown in Fig.~\ref{fig:Einfaches-Modell-Besch}
on the right, the difference of the thickness of the shell taken away
at the inner border of $\CM$ and added at the outer border decreases.
$\Delta a_{\CD}$ becomes more and more $\Delta a_{\CE}$, i.e. it
increases. Therefore, $a_{\CD}$ experiences accelerated expansion
as the growth $\Delta a_{\CD}$ in a fixed interval $\Delta t$ is
growing. From this explanation it becomes also clear that accelerated
expansion can only be a temporary effect. In Eq. (\ref{eq:acc-without-acc})
this is reflected by the fact that with the growth of the $\CE$--region
the parameter $\lambda_{\CM}$ is going to zero and with it $\ddot{a}_{\CD}$.
This explanation also shows that the effect depends on the number
of dimensions in which one is evaluating it. In a one--dimensional
setup, it will not appear at all as the two scale factors just add
linearly. It then grows with every dimension one adds what may be
seen in the growth of the second term in (\ref{eq:acc-split}). 
Adding generic volumes of the same shape but evolving differently, one has
geometric effects on the evolution of the global scale factor
(e.g. combine $4$ squares
to a bigger square but, as soon as you evolve the individual squares differently, the overall
shape will not be a square).

If we generalize the setup a little to include a decelerated expansion
of the $\CE$--region, Eq. (\ref{eq:acc-without-acc}) acquires an
extra $\ddot{a}_{\CE}$ term. If it is smaller than the $H_{\CE}^{2}$
term, one has acceleration of the whole configuration despite deceleration
of the inner constituent.

Note that the acceleration described above also shows up in a configuration
where the inner part is contracting, i.e. $H_{\CE}<0$. Then, the difference
between the shell that has to be taken away at the exterior of the
outer $\CM$--region and the one that has to be added on the inner
side of the $\CM$--shell will be growing. First, they are similar, but
finally, a reduction of the diameter of the $\CE$--region by $\Delta a_{\CE}$
will result in a smaller reduction of the overall diameter $a_{\CD}$,
i.e. $\Delta a_{\CD}<\Delta a_{\CE}$. This is to be interpreted
as acceleration of the $\CD$--region in the sense of shrinking deceleration.

To go one step further one could imagine to have a toy model of the
Universe build up out of balls of the first type. Each of them expands
in an accelerated way. If the accelerated phase sets in during a narrow
interval of time for all of the balls, one would find that the effect
adds up and can cause global acceleration. The setup in this extended
case is a bit similar to the Swiss cheese models, but without the
restriction of the embedding in an overall FLRW evolution.

\subsection{More general cases}

The emergence of accelerated volume expansion despite local deceleration
has been already widely discussed in the literature. Essentially it is due to the fact
that the local contributions are correlated in an average \cite{buchert:review}. 
To give a more intuitive explanation, R\"as\"anen, for example, argued in \cite{rasanen:de} that the physical reason for this volume acceleration
is that the volume fraction of the faster expanding regions rises.
In our partitioning approach this is reflected in Eq. (\ref{eq:acc-split}),
which may be recast in the form
\begin{equation}
\frac{\ddot{a}_{\CD}}{a_{\CD}}=\lambda_{\CM}\frac{\ddot{a}_{\CM}}{a_{\CM}}+\left(1-\lambda_{\CM}\right)\frac{\ddot{a}_{\CE}}{a_{\CE}}+\frac{2}{9}\left[\lambda_{\CM}\left(1-\lambda_{\CM}\right)\right]^{-1}\dot{\lambda}_{\CM}^{2}\;.
\end{equation}
It underlines the statement that there can be accelerated expansion
if the change of the volume fraction of the faster expanding regions
$\dot{\lambda}_{\CM}$ is sufficient. It also shows that the possibility
of acceleration holds beyond the toy model case. If structure formation
is rapid enough so that $\dot{\lambda}_{\CM}$ is able to counterbalance
$\ddot{a}_{\CM}/a_{\CM}$ and $\ddot{a}_{\CE}/a_{\CE}$, it will drive
acceleration.

From the averaged equations themselves one can derive the condition
for an acceleration of the volume scale factor. For strong backreaction
(See \cite{kolb:backgrounds} for the distinction between weak and
strong backreaction) Equation (\ref{eq:Raychaudhuri-Mittel}) tells
us that $\CQ_{\CD}$ must be positive and bigger than $4\pi G\average{\varrho}$
to lead to acceleration. In view of the definition (\ref{eq:Def-Q}) this
means that the term $\left(\average{\theta^{2}}-\average{\theta}^{2}\right)$
has to be sufficiently bigger than the shear term $\average{\sigma^{2}}$,
i.e. the Universe should be dominated by expansion
fluctuations rather than by fluctuations of the averaged rate of shear.

\section{Evaluation of the $N$--body simulation}

Here we want to describe how the results on the evolution of the $\lambda_{\CM}$
parameter in Figs. \ref{fig:Skalierung-von-Q} and \ref{fig:FLRW-Fit-Lambda}
were obtained. The data that we analyzed were obtained from a
simulation of the Virgo Supercomputing Consortium \cite{n-body}.
They trace the structure formation in a cube of a sidelength of $479h^{-1}$Mpc.
The underlying cosmological model was a $\Lambda$CDM model with $\Omega_{\Lambda}=0.7$
and $H_{0}=70\,\rm{km\, s^{-1}Mpc^{-1}}$. The simulation contained
$512^{3}\approx134\rm{Mio}$ particles.

\subsection{A simple mesh method\label{sub:A-simple-mesh}}

To obtain a rough approximation of the overdensity field, we separated
the data cube with a mesh with a fixed grid size and determined the
number of data points in the cells. Then we added up the volume of
the most dense cells until the sum of the points contained in the cells added
reached half the number of total points of the simulation volume.
This is because we decided to fix the characterization of the volumes
in the initial, near to homogeneous state. Under the assumption of
a Gaussian distribution of the density field with only small density
fluctuations, approximately one half of the mass in the Universe will
be in overdense $\CM$-- resp. underdense $\CE$--regions, if their typical
size was nearly the same at that epoch. The volume obtained by the
addition described was suspected to be $V_{\CM}$ and used to determine
$\lambda_{\CM}$ by dividing by the total box volume. Of course this
identification is a source of error because it may happen that in
a region that was underdense in the beginning structure formation
leads to a density peak that belongs to the densest half of the matter
distribution. But as the considerations in this paper are mostly concerned
with the influence of structure formation
on the global expansion history, we expect the error to be tolerable.

Apart from this problem the determination is well--defined even if
the $\CM$--regions scale differently than the global $a_{\Lambda\rm{CDM}}$--scaling
of the simulation. This is because we are only interested in volume
fractions and assume our background region $\CD$ to scale as $a_{\Lambda\rm{CDM}}$
in Section~\ref{sec:Modelling-structure-formation}, even if we assume
different physical reasons for this behavior. Then 
\begin{equation}
\lambda_{\CM}=\frac{V_{\CM}}{V_{\CD}}=\frac{N_{\CM}V_{box,initial}a_{\Lambda\rm{CDM}}^{3}}{N_{\CD}V_{box,initial}a_{\Lambda\rm{CDM}}^{3}}=\frac{N_{\CM}}{N_{\CD}}\;,
\end{equation}
where $N_{\CM}$ and $N_{\CD}$ are the number of boxes of the overdense
half and the total number of boxes respectively, and the simulation
volume just drops out.

\begin{figure*}
\includegraphics[width=0.47\textwidth]{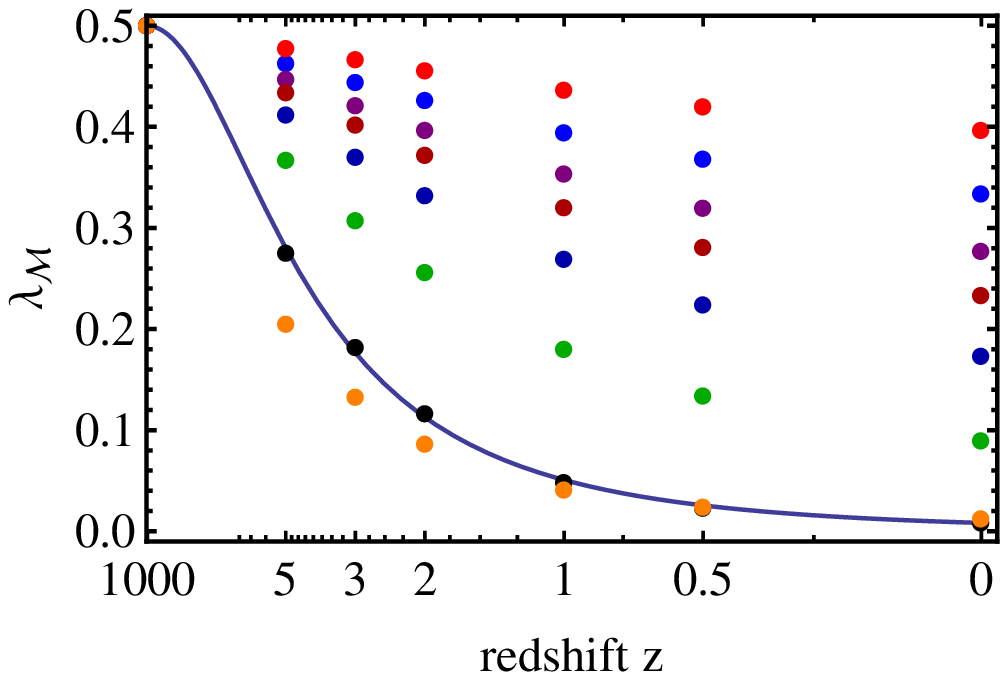}%
\hspace{0.05\textwidth}%
\includegraphics[width=0.47\textwidth]{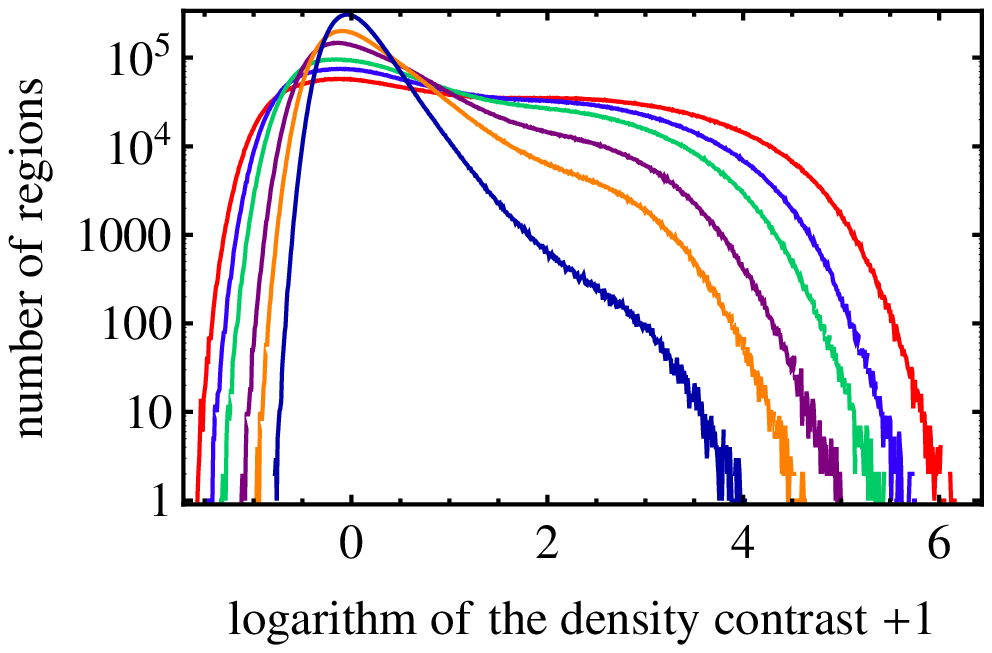}

\caption{Left: Comparison of the results for a box size of $R=1,\,5,\,10,\,15,\,20,\,30,\,50\,\rm{h^{-1}Mpc}$
(from bottom to top) with the one obtained by the Voronoi approach
(dots on the curve that represents the fit of Fig.~\ref{fig:FLRW-Fit-Lambda}).
The corresponding curve varies more strongly especially at the beginning
which may be interpreted as faster structure formation. Right: distribution
of the density contrast that has been assigned to the simulation points
by the Voronoi method. The lowest curve shows the initial distribution
of $z=5$. The other curves show the same distribution at redshifts
of $3$, $2$, $1$, $0.5$ and $0$ (from bottom to top). The values
on the $x-$axis represent $\log\left(\delta+1\right)$, the $y-$axis the
number of regions with the corresponding density contrast in a logarithmic
scaling.\label{fig:Lamda-Skala}\label{fig:Struktur-Statistik}}

\end{figure*}

To derive a more reliable estimate for $\lambda_{\CM}$ than the one
obtained by the procedure described above, one would have to consider
Lagrangian domains as required by the averaging formalism. This would
imply to have a clear definition of the volume associated to the particles
in the initial state and to follow exactly this volume throughout
the evolution. As in the real world, at least at small scales vorticity
is not absent,  
it would be complicated to put this into practice, especially in a relativistic framework. Other
complications include particle crossing and domain merging. We suppose, however, that the outcome
would not be completely different, since averaging strategies to find extensive quantities over smoothed--out singularities are implementable and would likely not change the partitioning.
It is clear that there is a huge potential to improve on our rough evaluation.

The above procedure was carried out for different grid sizes to get
an impression of the dependence of $\lambda_{\CM}$ on the grid length.
The result is shown on the left--hand side of Fig.~\ref{fig:Scaling}.
We used values of $1$--$50\rm{h^{-1}Mpc}$. The plot makes clear
that there is in fact a strong dependence of $\lambda_{\CM}$ on the
grid size. This is not surprising as structure formation in the Universe
proceeds in a hierarchical way. The small--scale structure forms first
and then starts to combine to larger structures. This is reflected
in the steeper decrease of $\lambda_{\CM}$ for smaller grid lengths.

To motivate the choices of grid resolution in Sec.~\ref{sec:A-simple-scaling}
we consider that a scale of $1\rm{h^{-1}Mpc}$, as used for the
lowest curve of Fig.~\ref{fig:Lamda-Skala}, may already be too small
as the simulation is used for large--scale structure. The highest curves
above $10\rm{h^{-1}Mpc}$ grid length is on the other hand already
merging typical overdense structures like clusters with underdense
voids. If one takes into account that a typical cluster of galaxies
is in the range of $2-8\rm{Mpc}$ one would expect that grid
sizes of $2-8\rm{h^{-1}Mpc}$ are perhaps best adopted to the
considered problem. Therefore, we chose a grid length of $5\rm{h^{-1}Mpc}$
for the comparison of Fig.~\ref{fig:Skalierung-von-Q}.

For the results of Section~\ref{sub:Quantitative-conclusions} we
had to find an adequate functional form for the evolution of $\lambda_{\CM}\left(a_{\CD}\right)$.
From the theoretical point of view it should meet several requirements.
First of all, we expect $\lambda_{\CM}$ to start at an initial value
of about $0.5$ on all scales, as the density fluctuations of the
Early Universe are assumed to be Gaussian. The second condition should
be that it does not deviate rapidly from this value at the beginning.
So $\lambda_{\CM}$ should have a flat tangent at $a_{\CD}=1$. Third,
we expect that $\lambda_{\CM}$ tends to $0$ on all scales as matter
clusters strongly and the voids keep expanding. For a standard $\Lambda$CDM
model this is even more true as then the cosmological constant will
accelerate the growth of the empty regions between the matter filaments.
These three conditions lead to the ansatz of (\ref{eq:Fit-komplex}).

\subsection{The Voronoi method and structure formation}

As the fixed mesh is not adapted to the real matter distribution in
the simulation volume, we used in a second step a partitioning of
the volume into Voronoi regions. This approach has been advocated
by van de Weygaert \cite{icke:voronoi,weygart} and assigns each simulated
point the region that is closer to it than to any other point. In
condensed matter physics a Voronoi cell is better known as the Wigner
Seitz cell. To calculate the Voronoi cells we used the program {}``qhull''
\cite{qhull}, and derived $\lambda_{\CM}$ in the same way as described
above by adding the densest Voronoi cells. The resulting points are
shown in Fig.~\ref{fig:Lamda-Skala} on the left together with the
curves for $0.5$, $1$ and $5\rm{h^{-1}Mpc}$, as derived by the
mesh method. The plot shows that the Voronoi results are closest to
the $1\rm{h^{-1}Mpc}$--mesh ones. The shape, however, shows a steeper
behavior in the beginning, which one would interpret as a faster structure
formation. The Voronoi result was used for the fit of $\lambda_{\CM}$
of Fig.~\ref{fig:FLRW-Fit-Lambda}.

We finally use the separation into Voronoi cells to show the formation
of structure in the distribution of the overdensity field.

To this end we plot in Fig.~\ref{fig:Struktur-Statistik} on the
right the evolution of the distribution of the density contrasts of
the Voronoi cells. For our first data at a redshift of $z=5$ the
distribution still is strongly peaked around zero. The maximum is
then moving to underdense regions. In the meantime a second maximum
emerges for high density regions. The mean overdensity, however, stays
zero all the time. It is interesting to see that the distribution
spreads and reaches an extension of $8$ orders of magnitude in the density
contrast. These extreme differences in the local densities and their
expected different evolution is one of the main motivation for the
idea of a structure formation effect on the global evolution.
\end{document}